\documentclass[aps,pra,floatfix,twocolumn,tightenlines,superscriptaddress,notitlepage]{revtex4-1}

\usepackage{graphicx}
\usepackage{amsmath}
\usepackage{latexsym}
\usepackage{graphicx}
\usepackage{multirow}
\usepackage{url}
\usepackage[bookmarks, colorlinks=true, pdftitle={Performance of a radiatively cooled system for quantum optomechanical quantum experiments in space},
            pdftex
            ]{hyperref}

\makeatother

\begin{document}

\title{Performance of a radiatively cooled system for quantum optomechanical
experiments in space}

\author{Andr{\'e} Pilan-Zanoni}
\email[Corresponding author: ]{plan.z@outlook.com}
\affiliation{Airbus Defence and Space GmbH, Friedrichshafen, Germany}

\author{Johannes Burkhardt}
\affiliation{Airbus Defence and Space GmbH, Friedrichshafen, Germany}

\author{Ulrich Johann}
\affiliation{Airbus Defence and Space GmbH, Friedrichshafen, Germany}

\author{Markus Aspelmeyer}
\affiliation{Vienna Center for Quantum Science and Technology, University of Vienna, Austria}

\author{Rainer Kaltenbaek}
\email[Corresponding author: ]{rainer.kaltenbaek@univie.ac.at}
\affiliation{Vienna Center for Quantum Science and Technology, University of Vienna, Austria}

\author{Gerald Hechenblaikner}
\altaffiliation[Present address: ]{European Southern Observatory (ESO), Garching bei M{\"u}nchen, Germany}
\affiliation{Airbus Defence and Space GmbH, Friedrichshafen, Germany}

\begin{abstract}
The performance of a radiatively cooled instrument is investigated
in the context of optomechanical quantum experiments, where the environment
of a macroscopic particle in a quantum-superposition has to be cooled
to less than 20\,K in deep space. A heat-transfer analysis between
the components of the instrument as well as a transfer-function analysis
on thermal oscillations induced by the spacecraft interior and by
dissipative sources is performed. The thermal behaviour of the instrument
in an orbit around a Lagrangian point and in a highly elliptical Earth
orbit is discussed. Finally, we investigate further possible design improvements aiming at
lower temperatures of the environment of the macroscopic particle. These include a mirror-based design of the imaging system on the optical bench and the extension of the heat shields.
\end{abstract}

\maketitle

\section{Introduction}
In order to perform quantum experiments with macroscopic systems,
they need to be well isolated from the environment. For optomechanical systems, a mechanical support can pose a limit for such isolation due to thermal and vibrational coupling to the environment. Even if such systems do not have a mechanical support but utilise optically or electromagnetically trapped test particles, the trap may have to be switched off to unobstructedly observe quantum effects. This implies that the test particles ideally are in free
fall \citep{RomeroIsart2011b,MAQROtechnical}. 

Earth-bound experiments are limited in this respect as the particle
is subject to short free-fall times. In contrast,
deep space offers favourable conditions due to microgravity and very
long free-fall times. In addition, deep space also offers an outstanding vacuum quality (ca. $10^{-15}\,$mbar) and low temperatures (ca. $2.7\,$K). Using these environmental benefits allows minimizing the coupling of the particle to the environment due to scattering of gas molecules and plasma particles as well as scattering and absorption of blackbody radiation. Optimally harnessing the low deep-space temperature requires a thermal-design concept to shield the quantum system from heat loads from the spacecraft and nearby celestial bodies.

Instead of utilizing the deep-space environment, an alternative is to use active cooling (see, e.g., Ref.~\citep{swinyard2009space}). This approach has several disadvantages. Active cooling requires cryogenic helium tanks and pumps. These add extra mass that has to be launched to space and requires more mission control variables. Because helium is a very evasive gas, it may interact with the test particle, leading to additional decoherence. Moreover, the duration of the experiment in deep space would be limited by the quantity of helium stored in the tanks.

These limitations can be overcome by using a passively cooled system taking direct advantage of the deep-space environment. Concepts of passive radiatively cooled systems were suggested in Ref.~\citep{TimHowardenPaper},
where a set of shields were employed to block sun radiation from reaching
telescope mirrors. In doing so, a temperature of 20\,K was achieved in Ref.~\citep{TimHowarden2}. Modern space missions such as James Webb \citep{JamesWebb},
Gaia \citep{de2012science,urgoiti2005mechanisms} and Herschel/Planck
\citep{pilbratt2010herschel,tauber2010planck} also make use of similar
passively cooled systems. In addition to their thermal function, those
shields can also act as wake shields \citep{oran1977preliminary}
and be used for protecting the test particle against solar wind and
spacecraft outgassing \citep{MAQROtechnical}.

Using a system of this type for quantum optomechanical experiments was first suggested in the context of the mission proposal MAQRO (macroscopic quantum resonators)\citep{MAQROtechnical} and investigated in more detail in Refs.~\citep{TN3,MAQROpaper1}. MAQRO aims at testing the foundations of quantum physics in a novel parameter range far beyond what is achievable in Earth-bound experiments. In particular, MAQRO proposes using long-lived macroscopic quantum superpositions of dielectric test particles with a mass up to about $10^{11}\,$atomic mass units (amu)\citep{MAQROtechnical}. The experiment is to be performed on an optical bench where the test particle is optically trapped in the intra-cavity field of a high-finesse cavity\citep{Chang2010a,RomeroIsart2010a,Barker2010a}. Additional cavity modes are used to cool the centre-of-mass (CM) motion of the particle. After this initial preparation, the test particle is released and propagates freely. Then the particle is prepared in a macroscopic superposition and again propagates freely until finally the position of the particle is measured. After many repetitions, the recorded positions should show an interference pattern. By repeating this experiment with particles of different radii and mass densities, one can investigate the parameter dependence of the interference visibility. This will allow for conclusive tests of a number of theoretical models predicting a quantum to classical transition for massive particles\citep{Bassi2013a}. Moreover, such experiments will allow quantitative tests of decoherence mechanisms due to interactions with the environment\citep{Hackermueller2004a,Hackermueller2003a}, and they will provide an experimental benchmark for future theoretical models that may predict deviations from quantum theory, e.g., models of quantum gravity.  

During the times of free evolution, the particle is in free fall while all optical fields are switched off. In order to achieve non-negligible interference visibility, all decoherence mechanisms during the free evolution of
the quantum superposition must be minimised. Given low ambient gas pressure, the dominant sources of decoherence apart from the emission of blackbody radiation are the scattering and absorption of blackbody radiation \citep{RomeroIsart2011b,RomeroIsart2011c,MAQROtechnical}. For that reason, an environment
temperature lower than $20\,$K has to be maintained throughout the experiment\citep{MAQROtechnical}.

An initial thermal-shield design proposed in \citep{MAQROtechnical,MAQROpaper1}
focused on determining the lowest temperature achievable by means
of passive cooling in space. In that work, a set of shields was optimised
in terms of number of shields, shield opening angles and shield distances
in order to minimise the temperature of a small test volume representing
the immediate environment around the nanoparticle. The results showed
that a steady-state temperature of $16.3\,$K can be reached including
anticipated optical and electrical dissipation on the optical bench
as well as heat transfer from the spacecraft via conduction and radiation.
However, performance issues such as heat-transfer dynamics and orbital
cases remained to be investigated. 

\begin{figure*}[htbp]
 \begin{center}
  \includegraphics[width=0.9\linewidth]{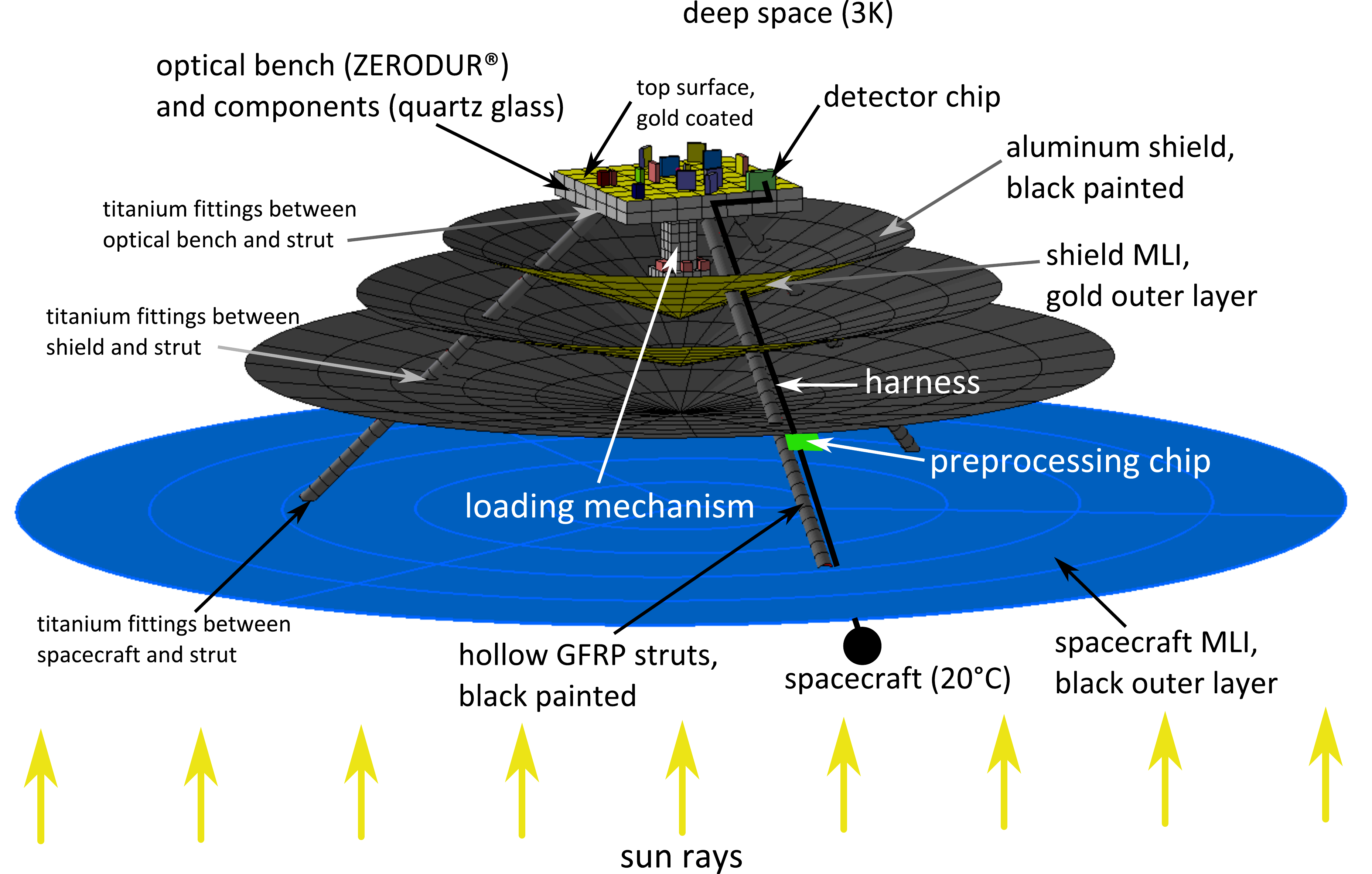}
  \caption{Schematic representation of the geometric mathematical model of the MAQRO instrument. \label{fig:GMM_MAQRO}.}
  \end{center}
\end{figure*}

Here, we aim at optimizing the thermal-shield design for achieving
even lower temperatures via passive cooling for quantum optomechanical
experiments in space. For that purpose, we conducted a heat-flow analysis
of the entire system as well as a transfer-function analysis. The
latter is used for evaluating the attenuation of the thermal fluctuations
from their origin in spacecraft all the way to the region where the
experiment is performed. Moreover, we improved the design of the optical
bench by replacing refractive optical elements with reflective ones.

\section{Modeling approach}
We model the scientific instrument for our thermal analysis using
the software ESATAN-TMS \citep{ESATANmanual}. The values of the radiative
couplings between surface nodes of the instrument are calculated with
the aid of a geometric mathematical model (GMM) using the Monte Carlo
ray tracing method. These values are fed into a thermal mathematical
model (TMM), in which further inputs for solution of the energy equation
are included. These inputs are the conductive couplings between instrument
nodes, internal and external heat loads, boundary conditions and parameters
for numerical processing. The TMM makes use of the lumped-parameter
formulation, where all properties of a node are concentrated in
its barycenter.

\subsection{The optical bench and the geometric mathematical model}
\begin{figure}[htbp]
 \begin{center}
  \includegraphics[width=0.97\linewidth]{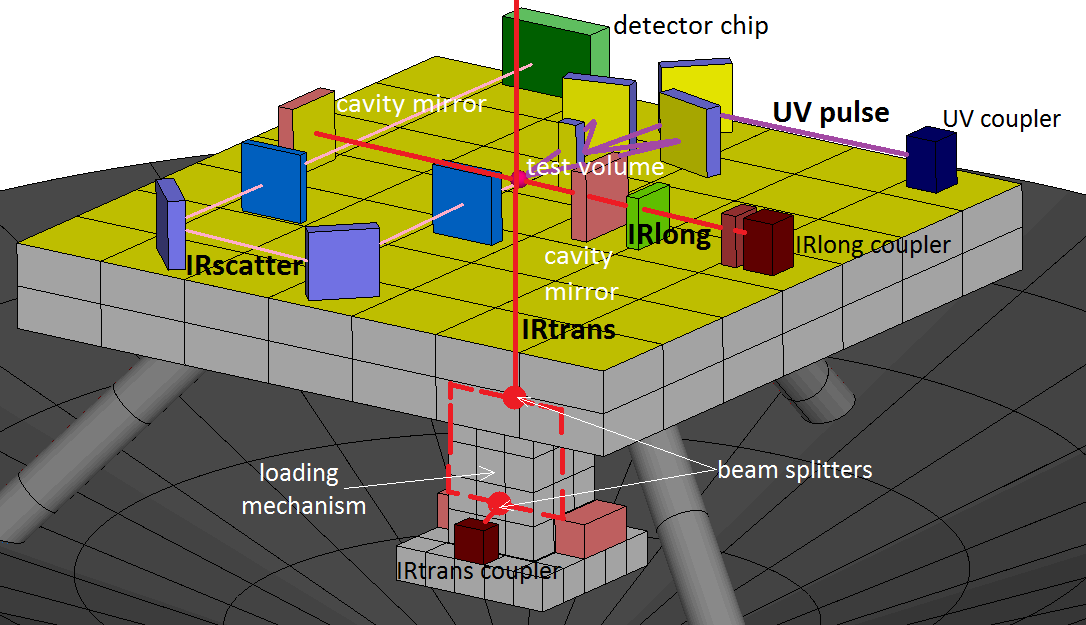}
  \caption{Schematic representation of the optical fields on the optical bench \label{fig:Optical-fields-bench}.}
  \end{center}
\end{figure}

Figure \ref{fig:GMM_MAQRO} schematically illustrates the geometric
mathematical model for a radiatively cooled instrument of MAQRO \citep{MAQROtechnical,TN3}
as it was first analysed in \citep{MAQROpaper1}. Figure \ref{fig:Optical-fields-bench}
shows the optical bench and the optical fields.

The disc at the bottom of Figure~\ref{fig:GMM_MAQRO} represents the
spacecraft MLI with an outer layer of black Kapton. 

In the quantum experiments proposed in MAQRO, the central element is a dielectric nanosphere. Various methods for 
loading such particles into an optical trap in ultra-high-vacuum (UHV) 
conditions are being investigated. In Figure \ref{fig:GMM_MAQRO}, this is
referred to as loading mechanism. From the loading region, the particle is propelled towards the cavity mode via radiation pressure by an IR beam. Two additional IR beams, represented as IRtrans in Figure
\ref{fig:Optical-fields-bench}, are used to confine the region where
the particle should be during the experiment. If the particle leaves
that region, it scatters light that can be detected by a chip located
on the optical bench, here called detector chip \citep{Chip2}. In
this case, the particle has to be transported back into the region
of interest along the cavity mode. The immediate
environment of the position where the nanosphere is supposed to be
located is referred to as test volume\citep{MAQROpaper1}. This test volume is modelled
as a black body with $\alpha=\varepsilon=1$ and zero specific heat
capacity ($c=0$).

The central element on the optical bench is an asymmetric optical
cavity with a cavity length of 97\,mm consisting of two high-reflectivity
spherical mirrors with radii of curvature of 75\,mm and 30\,mm.
Two optical modes of IR light (1064\,nm), represented as IRlong in
Figure \ref{fig:Optical-fields-bench}, are fed into the cavity from
an IR fibre coupler. The two modes have orthogonal polarisation and
are shifted in frequency with respect to each other by one free spectral
range (FSR) of the cavity. One mode is used to trap the particle.
The second mode can be used to either read out the particle position
or to cavity-cool the center-of-mass motion of the trapped particle
\citep{Lasercooling3,Lasercooling5,Lasercooling1}. The beam waist
size of the cavity mode is $64\,\mathrm{\mu m}$.

A combination of mirrors and lenses is used to image the trapping
region onto a detector chip and to detect the light scattered from
the nanosphere (IRscatter).

In addition to these optical elements for the cavity modes and the
scattered light, the optical design also contains several reflective
optical elements for UV light. These are intended to illustrate a
possible way for the preparation of a macroscopic superposition via
applying a short and well-focused UV beam to the trapping region.
If light is scattered by the nanosphere, its wavefunction will be
well localised. If no light is scattered, the particle will be in
a quantum superposition of being located anywhere outside the region
illuminated by the UV beam \citep{MAQROtechnical,Kaltenbaek2013}. Alternative 
methods of preparing the quantum superposition are currently investigated.

Except for the optical dissipation of the cavity mirrors, our thermal
simulations assume the optical fields to be turned off as in the case
during wavefunction evolution between preparing and detecting the
quantum state. Note, that this is a worst case estimation because during 
the time of wave-function expansion, strong sources of dissipation like the 
CMOS and preprocessing chips may actually be turned off.  

For our simulation, we assumed the optical bench as well as the structural
elements of the loading mechanism to be made of ZERODUR\textregistered{},
which is also used for the optical design in the LISA and LISA Pathfinder
missions \citep{LISAarticle1}. The upper surface of the bench is
coated with gold in order to minimise radiative heat transfer to the
test volume \citep{MAQROpaper1}. We assume the material of all optical
elements on the optical bench to be fused silica and that these elements
are bonded to the bench through hydroxide-catalysis bonding described
in \citep{Bonding2005}. 

The struts support the optical bench and connect it to the spacecraft.
Each of the three struts is made of a hollow GFRP tube filled with
polyurethane foam in order to avoid radiative heat exchange between
the internal walls of the strut \citep{MAQROpaper1}. A set of shields
are modelled in order to prevent spacecraft heat radiation from reaching
the bench as also seen in \citep{TimHowardenPaper}. These shields
consist of an aluminium plate whose bottom side is covered with MLI
consisting of 20 layers. The aluminium plate is black painted on its
top side and gold coated on its lower side. The MLI is aluminised
on its side to the shield and coated with a thin gold layer on its
side to the spacecraft. The opening angles and relative distances
are chosen to have the optimum values derived in Ref.~\citep{MAQROpaper1}.

The boundary nodes for our simulation are deep space at 3\,K and
the spacecraft. For the constant internal temperature of the spacecraft
we assume $20\,^\circ$C. This value is representative for the
constant internal temperature seen in various payloads. Components
such as harness, optical fibres and the preprocessing chip were not
geometrically modelled due to their small emitting areas, which makes
their radiative influence on the bench temperatures negligible \citep{MAQROpaper1}. 

For the radiative analysis, we used the Monte Carlo ray-tracing method to 
determine the radiative exchange factors between
the different surfaces of the geometrical model. In the analysis of
the radiative model, we use a fixed quantity of $100,000\,$rays. These are
propagated from various emitting surface elements to the test volume.
We sample a fixed quantity of 10\,000 rays propagating towards other
instrument components. The radiative coupling $GR$ between two model
nodes $i$ and $j$ is calculated using equation \ref{eq:GR_calc}
as described in \citep{ESATANmanual}.

\begin{equation}
GR(i,j)=A_{i}\varepsilon_{i}\alpha_{j}F_{ij}\label{eq:GR_calc}
\end{equation}

The Monte Carlo ray-tracing method is also used for the orbital cases
to evaluate external heat loads on the instrument such as solar, albedo
and Earth infrared heat \citep{ESATANmanual}.

\subsection{The thermal mathematical model and the energy equation}
The conductive couplings $GL$ between two nodes in the TMM can be calculated
using equation \ref{eq:GL_calc} (see also Ref.~\citep{ESATANmanual}).

\begin{equation}
GL(i,j)=\frac{k(T)S_{ij}}{d_{ij}}\label{eq:GL_calc}
\end{equation}

The preprocessing chip, as described for instance in \citep{Loose2005a},
is fixed below the first shield so that its dissipation does not directly
affect the optical bench. The harness that connects the detector chip,
the preprocessing chip and the spacecraft as Figure \ref{fig:GMM_MAQRO}
schematically shows is assumed to consist of steel. For simplicity, the thermal properties
of the chips are assumed to be the same as for quartz glass.

Based on the results of Ref.~\citep{MAQROpaper1}, we neglected 
optical fibres in our model. They have little influence on the 
bench temperature due to their low heat conductivity. The spacecraft, 
the shields and the optical bench are assumed to be connected to the 
struts using titanium fittings. These are included in the thermal model as constant conductive couplings \citep{MAQROpaper1}.

The rate of heat flow between two different nodes $i$ and $j$ can
be evaluated using the following equations for the radiative heat
flow $\dot{Q}_{\mathrm{R},ij}$ and the conductive heat flow $\dot{Q}_{\mathrm{L},ij}$, respectively:

\begin{equation}
\dot{Q}_{\mathrm{R,\mathit{ij}}}=GR(i,j)\sigma\left(T_{i}^{4}-T_{j}^{4}\right)\label{eq:QGR_calc}
\end{equation}

\begin{equation}
\dot{Q}_{\mathrm{L},ij}=GL(i,j)\left(T_{i}-T_{j}\right)\label{eq:QGL_calc}
\end{equation}

For each single node $i$, the energy equation can be evaluated using
equation \ref{eq:energy_balance_equation} (see Ref.~\citep{ESATANmanual,Fasoulas2011}):
\begin{equation}
\resizebox{0.8\linewidth}{!}{$\sum_i \left( \dot{Q}_{\mathrm{R\mathit{,i}}}+\dot{Q}_{\mathrm{L\mathit{,i}}}+\dot{Q}_{\mathrm{I\mathit{,i}}}+\dot{Q}_{\mathrm{S\mathit{,i}}}+\dot{Q}_{\mathrm{E\mathit{,i}}}\right)=m_{i}c_{i}\frac{\partial T_{i}(t)}{\partial t}$}\label{eq:energy_balance_equation}
\end{equation}
Here, the indices R, L and I refer to radiative, conductive and dissipation
heat, respectively. S refers to solar heat, E to Earth albedo and
infrared radiation from the Earth.

The dissipations considered in the TMM are: $10.0\,$mW for the preprocessing
chip, $1.0\,$mW for the detector chip and $0.2\,$mW of the cavity mirrors
\citep{MAQROpaper1}. In our simulations, we assumed that all these sources of dissipation are continuously active. This represents a worst-case scenario because the dissipation will, in fact, be significantly less during the long free-fall times of the test particle~\citep{MAQROtechnical}.

The energy equation is evaluated for each node of the model and treated
numerically as shown in \citep{ESATANmanual}. An iterative and an
inverse matrix solver are used for the steady state. As a plausibility 
check for our analysis, we compared the temperature predicted by these
two methods for the test volume. The difference was at most $0.1\,$K.
For the transient cases, we used the Crank-Nicolson method\cite{Crank1947a} and the backward
differentiation formula based on the Gear formalism\cite{Shampine1979a}.

\subsection{Transfer-function analysis}
The amount of solar radiation the spacecraft experiences varies in the course
of HEO orbit. This can lead to small oscillations of the internal temperature
of the spacecraft. We can investigate the influence of such oscillations on the
instrument temperature using a transfer-function analysis in steady state.
For that purpose, the energy equation is linearised and transformed
through La\-pla\-ce transformation from the time to the frequency
domain as described for instance in \citep{altenburg2008application}.
Using this approach, we can calculate the relation between the temperature oscillations of the spacecraft and resulting changes in temperature at various instrument components. Due to the linearisation, only small temperature variations of the spacecraft can be considered.

This relation, also referred to as gain, is calculated with the help
of the software TransFAST developed by the company Airbus Defence \& Space (ADS, formerly
Astrium GmbH) \citep{TransFAST2010}.

\section{Results and discussion}
\begin{figure}[htbp]
 \begin{center}
  \includegraphics[width=0.97\linewidth]{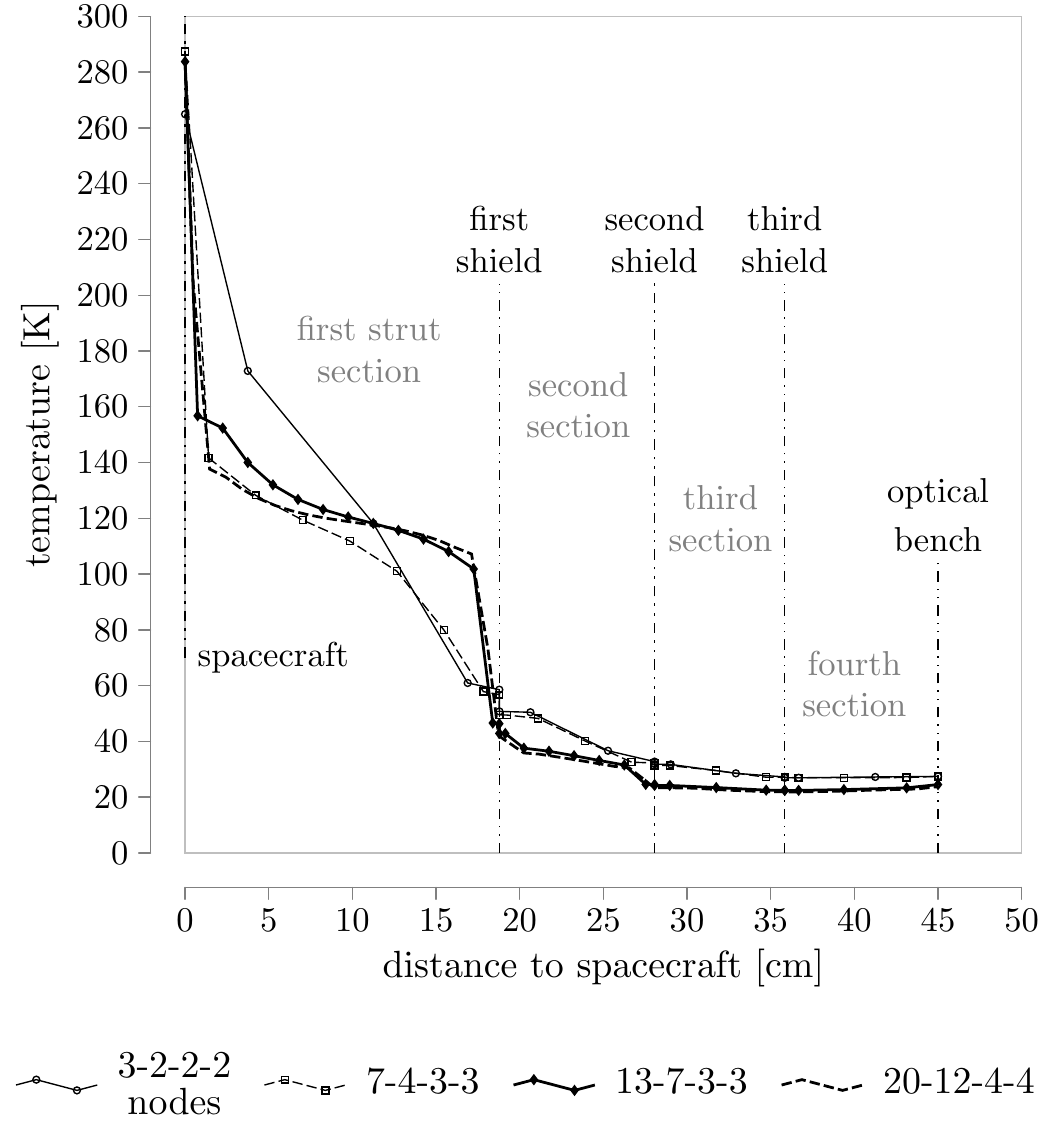}
  \caption{Temperature distribution along an individual strut for three different discretisation approaches \citep{PilanZanoni2014}.\label{fig:Temperature-strut-1}}
  \end{center}
\end{figure}

Because the temperature difference between spacecraft
and the optical bench is around 266\,K, the struts needed to be highly
discretised. Figure \ref{fig:Temperature-strut-1} shows the temperature
distribution along an individual strut depending on the node mesh. 

In our thermal analysis, we optimised our mesh to 13 axial nodes for the
first strut section, 7 axial nodes for the second section and 3 nodes for
the third section. Using even finer meshes is unnecessary because it
results in the same temperature distribution for the regions close to the
optical bench, as apparent from Fig.~\ref{fig:Temperature-strut-1}

It can be seen that 13 axial nodes for the first strut section, 7
axial nodes for the second and 3 nodes for the third and fourth sections
are adequate for detailing the temperature attenuation along the strut.

It can be shown that the maximum temperature difference along the surface of each aluminium shield was about $0.2\,$K and, along the surface of the optical bench, it was about $0.3\,$K\citep{PilanZanoni2014}. For this reason, we can assume the shields and the optical bench to have uniformly distributed temperatures. The same analysis showed that the test volume has a steady-state temperature of about $13.9\,$K and the optical bench about 24.7\,K. Due to the
minimisation of numerical averaging effects, a reduction in the temperature
of the test volume of about 2.4\,K is observed in comparison with
the results obtained in \citep{MAQROpaper1}.

\begin{figure*}[htbp]
 \begin{center}
  \includegraphics[width=0.9\linewidth]{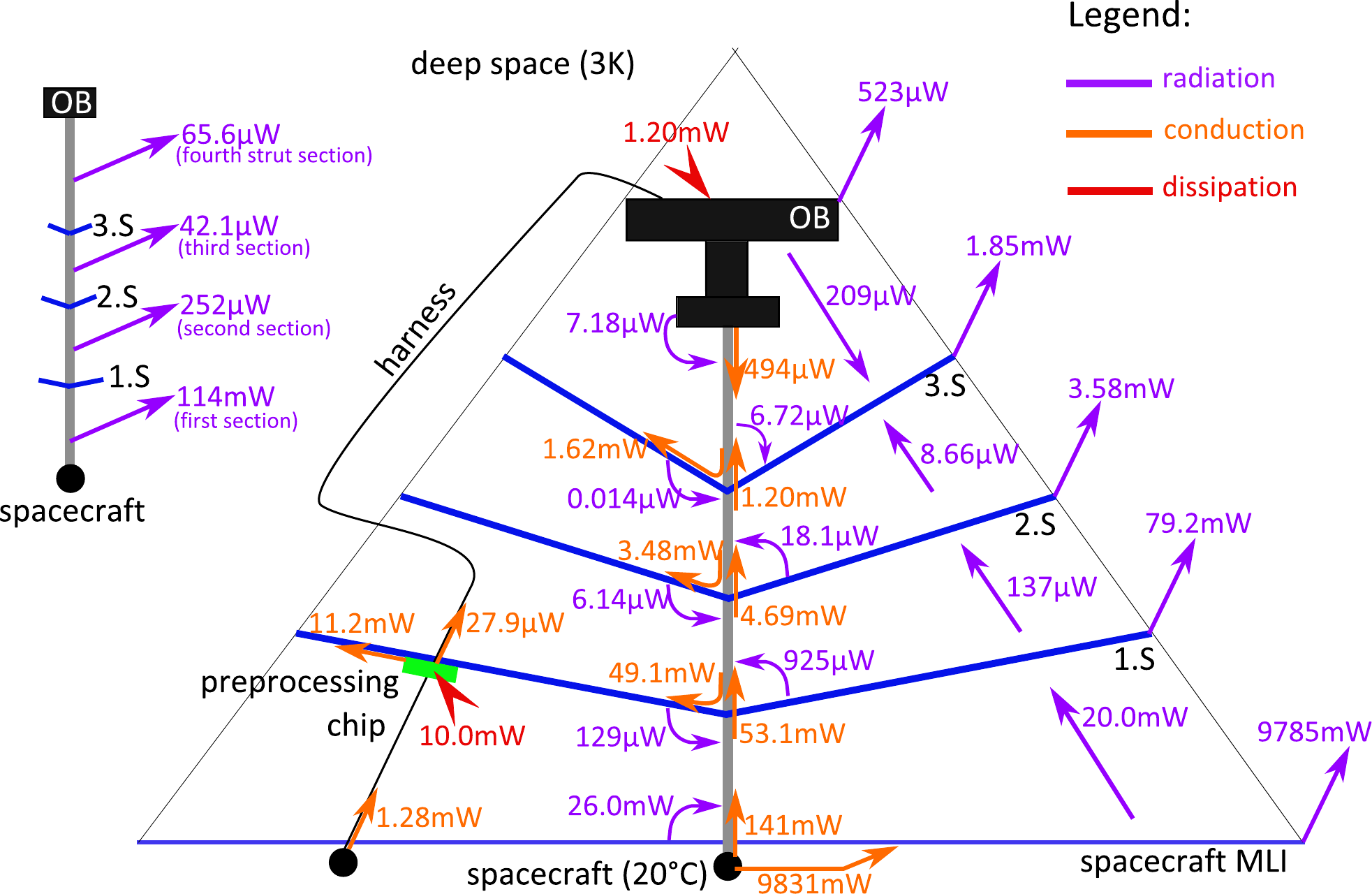}
  \caption{Schematic heat-flow diagram of the entire instrument \citep{PilanZanoni2014}.\label{fig:Schematic-heat-flux}}
  \end{center}
\end{figure*}

\subsection{Heat-flow analysis \label{sub:Heat-flux-analysis}}
We performed a heat-flow analysis for the entire MAQRO instrument
in steady state in order to quantify the heat transfer between the
different instrument components and to better assess the temperature
results. Figure \ref{fig:Schematic-heat-flux} describes the results
of this analysis.

The heat flow in the instrument is partially due to the temperature
difference between the spacecraft ($20\,^\circ$C) and deep space
(3\,K) and partially due to the dissipation sources represented as
red arrows in Figure \ref{fig:Schematic-heat-flux}. These sources
are divided into the dissipation of the preprocessing chip and the
dissipation of the bench, which comprises the dissipations of the
detector chip and of the cavity mirrors. Radiative heat flow is represented
in purple, conductive heat flow in orange.

The harness is represented as a loose black line in Figure \ref{fig:Schematic-heat-flux}.
The schematic representation shows the three shields (1.S, 2.S and
3.S), the optical bench (OB) and the spacecraft MLI (S/C MLI). For
the sake of simplicity, the three struts are represented as a single
gray tube and the radiative flow of each strut section to deep space
is detailed in a separate illustration on the left-hand side of Figure
\ref{fig:Schematic-heat-flux}. The thin black lines form a triangle
illustrating the line of sight between the spacecraft and the optical
bench. The entire optical bench must be inside that triangle in order
to avoid receiving direct radiation from the spacecraft MLI.

The dominant part (9785\,mW) of the heat energy of the spacecraft
surface is radiated to deep space directly from the spacecraft MLI.
To a smaller extent, energy is radiated to deep space from the shields.
Still, this plays an important role for the passive cooling system.
Firstly, the shields prevent the massive amount of energy from the
spacecraft from reaching the optical bench by blocking the direct line of sight between bench and spacecraft. Secondly, they act as radiators as they receive
energy through conduction from the struts and emit this heat to deep space via radiation.

Figure \ref{fig:Schematic-heat-flux} also demonstrates the advantage
of positioning the preprocessing chip below the first shield. The
10\,mW dissipation is conducted to the first shield and is emitted from there to deep space. This prevents this dissipation heat from
reaching the optical bench. Nevertheless, a very small part of it
still flows through the harness. This assumption is based on an idealised
connection between the preprocessing chip and the first shield, which
led to a conductive coupling of around $2.8\,\mathrm{W/K}$. For comparison,
the coupling between harness and preprocessing chip is $3.3\times10^{-5}\,\mathrm{W/K}$.

A sensitivity analysis showed that the temperature of the test volume
increases to around 20\,K if the dissipation of the detector chip
increases from 1\,mW to 12\,mW \citep{PilanZanoni2014}.

The dissipation of electronic and optical elements on the optical
bench results in the third shield having a lower temperature (22.4\,K)
than the optical bench itself (24.7\,K). Although virtually no radiative
heat from the other shields reaches the third one, this shield has
an important function in cooling the optical bench by receiving part
of the heat from the bench and radiating it to deep space. Therefore,
the removal of the third shield would cause a considerable increase
of the temperature of the test volume as observed in \citep{MAQROpaper1}
for a configuration with only two shields.

The first strut section removes a great part of the energy radiatively
to deep space. However, the amount of radiated heat from the struts
to deep space diminishes along the strut because of the thermal attenuation
induced by the shields. Note that, as consequence of the dissipation
on the optical bench, the fourth section radiates slightly more heat
to deep space than the third section.

Figure \ref{fig:Schematic-heat-flux} illustrates that the configuration
of the shields is well-optimised for passively cooling the optical
bench because the heat flow on the path from the spacecraft to the
optical bench is strongly attenuated. While increasing the number
of shields would yield only a very slight increase in that attenuation
\citep{MAQROpaper1}, it would significantly increase the complexity
of the instrument.

One can also conclude from the heat-flow diagram that the heat flow
in the instrument is consistent with the temperatures shown in Figure
\ref{fig:Temperature-strut-1}. For instance, the low temperature
obtained for the test volume and the optical bench is due to the fact
that the amount of heat that circulates in the bench is much smaller
than the amount that circulates in the shields and the struts.

\subsection{Transfer-function analysis \label{sub:Thermal-oscillations}}

\begin{figure}[htbp]
 \begin{center}
  \includegraphics[width=0.97\linewidth]{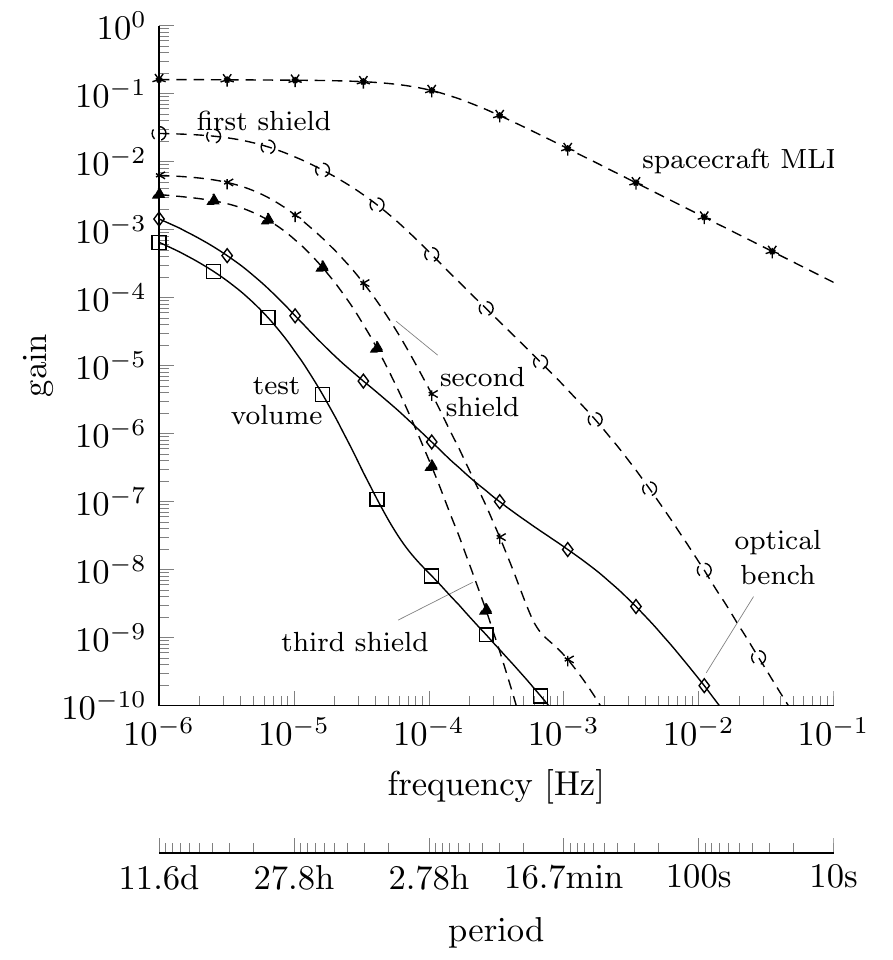}
  \caption{Results of the transfer-function analysis, plotted on a logarithmic scale for each of the major instrument
components, with the spacecraft temperature at $20\,^\circ$C as a single input.\label{fig:Transferanalyse_SC-1-1}}
  \end{center}
\end{figure}

We performed a transfer-function analysis in order to investigate
the effect of temperature oscillations of the spacecraft on the temperatures
of the instrument. We considered temperature variations with oscillation
frequencies in the spectral range between $10^{-6}$ to $10^{-1}\,\mathrm{Hz}$.
This corresponds to an oscillation period between 10\,s and 11.6
days, respectively. These periods cover the time scale of the optomechanical
experiments considered for MAQRO. For such measurements, the temperature
of the optical bench should be kept constant over a period of 100\,s
for a single measurement and from five to ten days for a full measurement
series \citep{MAQROtechnical}. Figure \ref{fig:Transferanalyse_SC-1-1}
shows the calculated gains of each instrument component. 

The gain is defined as the differential temperature at the output,
which can be any instrument component under test, divided by the differential
temperature at the input, such as fluctuations in the spacecraft.
Thermal disturbances are transferred on a much slower time scale compared,
for example, to electrical oscillations. Consequently, the inertia
of thermal fluctuations, especially at low frequencies, lead to larger
thermal gains. 

\begin{figure}[htbp]
 \begin{center}
  \includegraphics[width=0.97\linewidth]{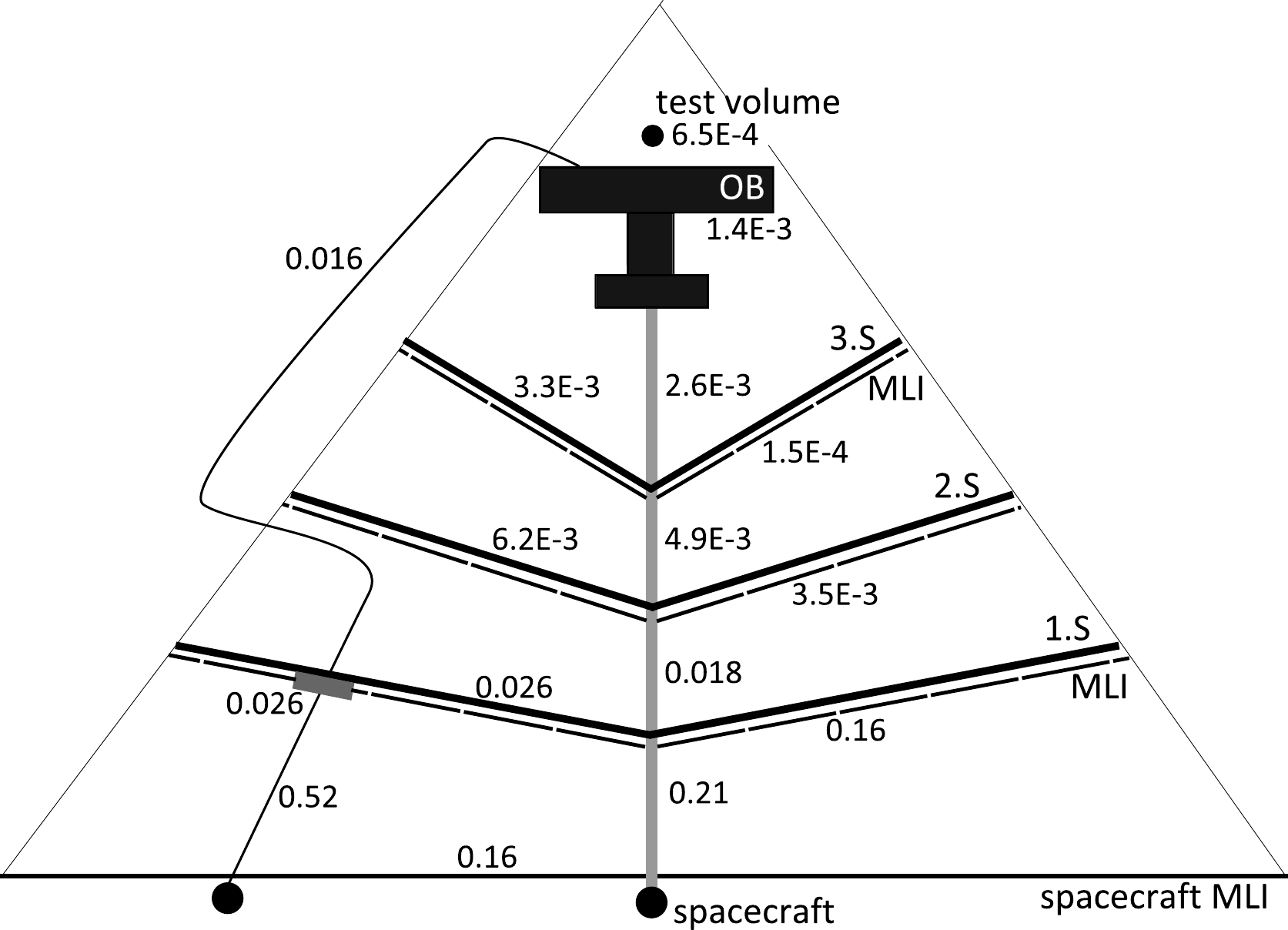}
  \caption{Calculated gains for different instrument components in {[}K/K{]} defined as the temperature response at the outputs divided by the input temperature oscillation. The spacecraft temperature at $20\,^\circ$C is the single input and oscillates at a frequency of $10^{-6}\,\mathrm{Hz}$ (period: $11.6\,$days). \label{fig:Tannenbaum_Transferanalyse_SC}}
  \end{center}
\end{figure}

Figure \ref{fig:Tannenbaum_Transferanalyse_SC} shows a detailed analysis
of gains across the instrument for a low frequency of $10^{-6}$\,Hz.
We chose to depict this analysis for that frequency because it induces
the largest gains. The gain for each strut section corresponds to
the average of the three struts belonging to the same section. 

Instrument components that are conductively linked to the spacecraft,
like the first section of the struts and the first section of the
harness, have the highest gains. For components radiatively coupled
to the spacecraft, the gains are smaller. Clearly, components that
are directly connected with the source of oscillation, like the spacecraft
MLI and the MLI of the first shield, are strongly influenced by the
temperature variations. This results in higher gains than for components
far from the source of oscillation.

We can conclude from the results shown in Figure \ref{fig:Tannenbaum_Transferanalyse_SC}
that temperature oscillations of the spacecraft have only very little
effect on the temperature of the optical bench. For the optical bench,
the gain is $1.4\times10^{-3}$. That means, a temperature change
of the spacecraft of 5\,K over 11.6 days only results in a change
of less than 0.01\,K in the temperature of the optical bench. Still,
even small temperature variations can have consequences on the fractional
frequency instability of the cavity on the optical bench of MAQRO.
We define this frequency instability as $\left|\Delta\nu/\nu_{\mathrm{Res}}\right|=\left|\Delta L/L\right|$.
Consider that the coefficient of thermal expansion (CTE) for ZERODUR\textregistered{}
is around $-0.63\times10^{-6}\,\mathrm{K^{-1}}$ at 30\,K, that the
cavity length $L$ is about 100\,mm and that the temperature fluctuations
are attenuated by a factor of $5\times10^{-10}$ for an oscillation
of 5\,K over a period of 100\,s. In this case, the relative change
in cavity length will be $\left|\Delta L/L\right|\approx 1.6\times10^{-16}$.
This is comparable to the best fractional frequency instabilities
achievable in ground-based cavity designs today, where the cavity
is isolated from temperature fluctuations outside the vacuum system
as well as possible \citep{numata2004thermal,notcutt2006contribution,amairi2013reducing}.

Because of the fact that the radiative terms are linearised for the
solution of the energy equation, the results obtained for the temperature
gains are only accurate for small variations. Nevertheless, it was
shown with a sensitivity analysis performed in \citep{PilanZanoni2014}
with the physically representative non-linear thermal model that the
temperatures of the test volume and of the optical bench remain small
for a wide range of input temperatures. Remarkably, even large changes of
the spacecraft temperature only lead to small changes in the temperature of
the test volume. For example, an increase of the temperature, say, from 
$80\,$K to $373\,$K induces only a temperature
increase in the optical bench from $22.8\,$K to $24.9\,$K and in the
test volume from $10.6\,$K to $11.6\,$K. Some of the input temperatures
are not applicable to a real spacecraft and were only analysed for
the purpose of the sensitivity analysis.

\begin{figure}[htbp]
 \begin{center}
  \includegraphics[width=0.97\linewidth]{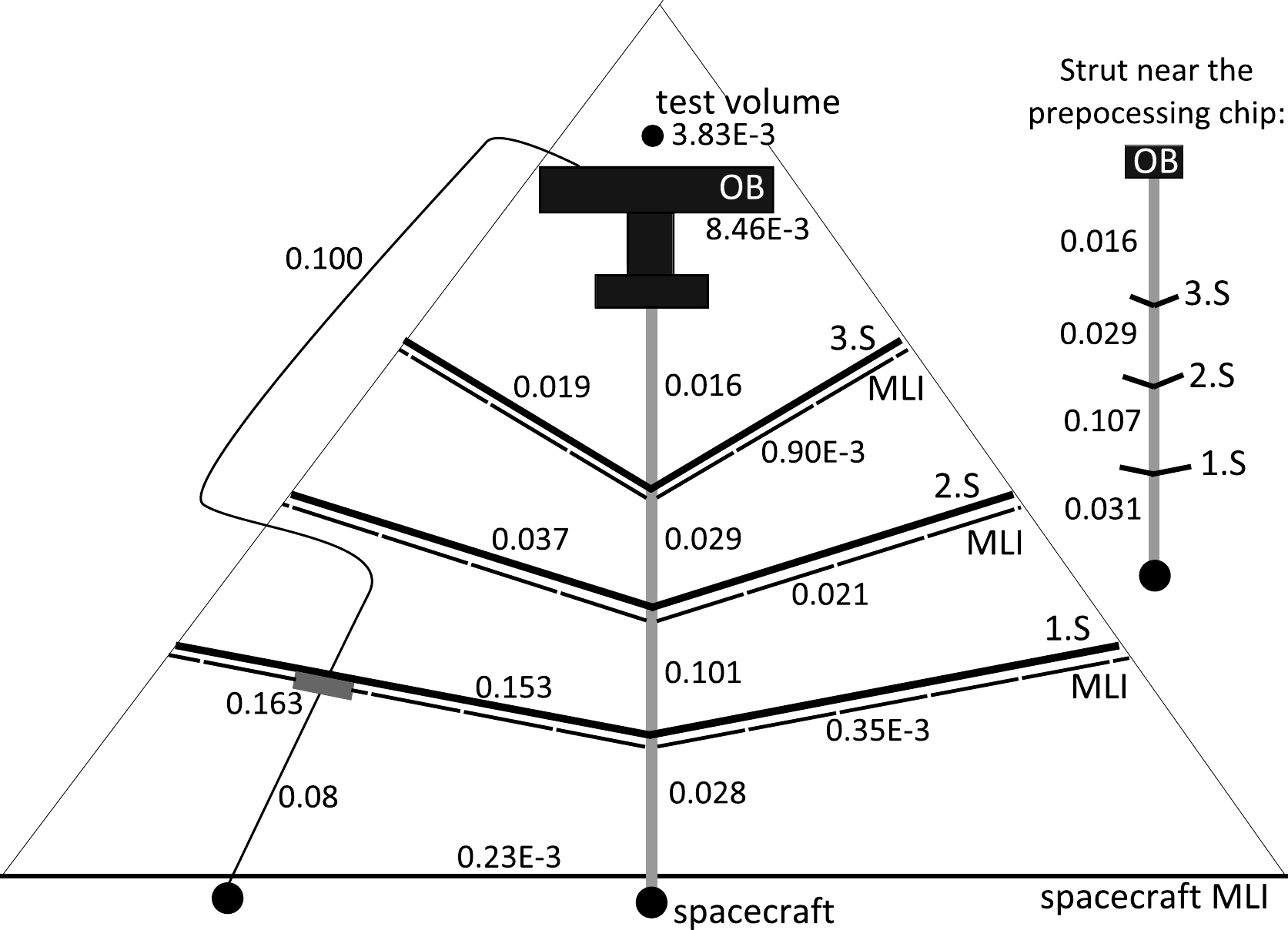}
  \caption{Calculated gains for different instrument components in {[}K/mW{]} defined as the temperature response
at the outputs divided by the input power oscillation. The $10\,$mW dissipation of the preprocessing chip is the single input and oscillates at a frequency of $10^{-6}\,\mathrm{Hz}$ (period: $11.6\,$days). On the right-hand side, a separate presentation of the gains of the strut near which the preprocessing chip is fixed. \label{fig:Tannenbaum_Transferanalyse_CCD}}
  \end{center}
\end{figure}

We also investigated the effect of oscillations of the dissipation
in the preprocessing chip. These may, for example, result from operating
a detector chip having a CCD sensor at varying frame rates. The gains
{[}in K/mW{]} are now defined as the differential temperature at different
components {[}in K{]} divided by the differential power at the input
{[}in mW{]}, which is now the dissipation of the preprocessing chip.
These gains are presented in Figure \ref{fig:Tannenbaum_Transferanalyse_CCD}.
Although electric components exhibit much higher oscillation frequencies
in reality, this investigation is restrained to a very low frequency
of $10^{-6}$\,Hz which represents a worst-case scenario for thermal
oscillations. 

The aluminium plate of the first shield and both harness sections present
the highest gains because they are directly connected to the preprocessing
chip via thermal conduction. The MLI of the first shield has a very
low gain because it is neither coupled conductively nor radiatively
with the preprocessing chip.

Because of the fact that the preprocessing chip is mount\-ed below
the first shield near one of the struts, the gain of this strut is
higher than the gains of the other ones. However, the gains of the
struts from the third section on are approximately the same as the
dissipation of the preprocessing chip is less dominant. 

The heat oscillation of the preprocessing chip at a frequency of $10^{-6}$\,Hz
induces a temperature variation of the optical bench of around $8.5\times10^{-3}\,\mathrm{K}/\mathrm{mW}$.
As a result, the temperature of the optical bench would increase by
less than 0.01\,K if the dissipation increased by 1\,mW. This leads
to a frequency instability of the optical cavity similar to that derived
in the case of fluctuations of the spacecraft temperature.

\begin{figure*}[htbp]
 \begin{center}
  \includegraphics[width=0.45\linewidth]{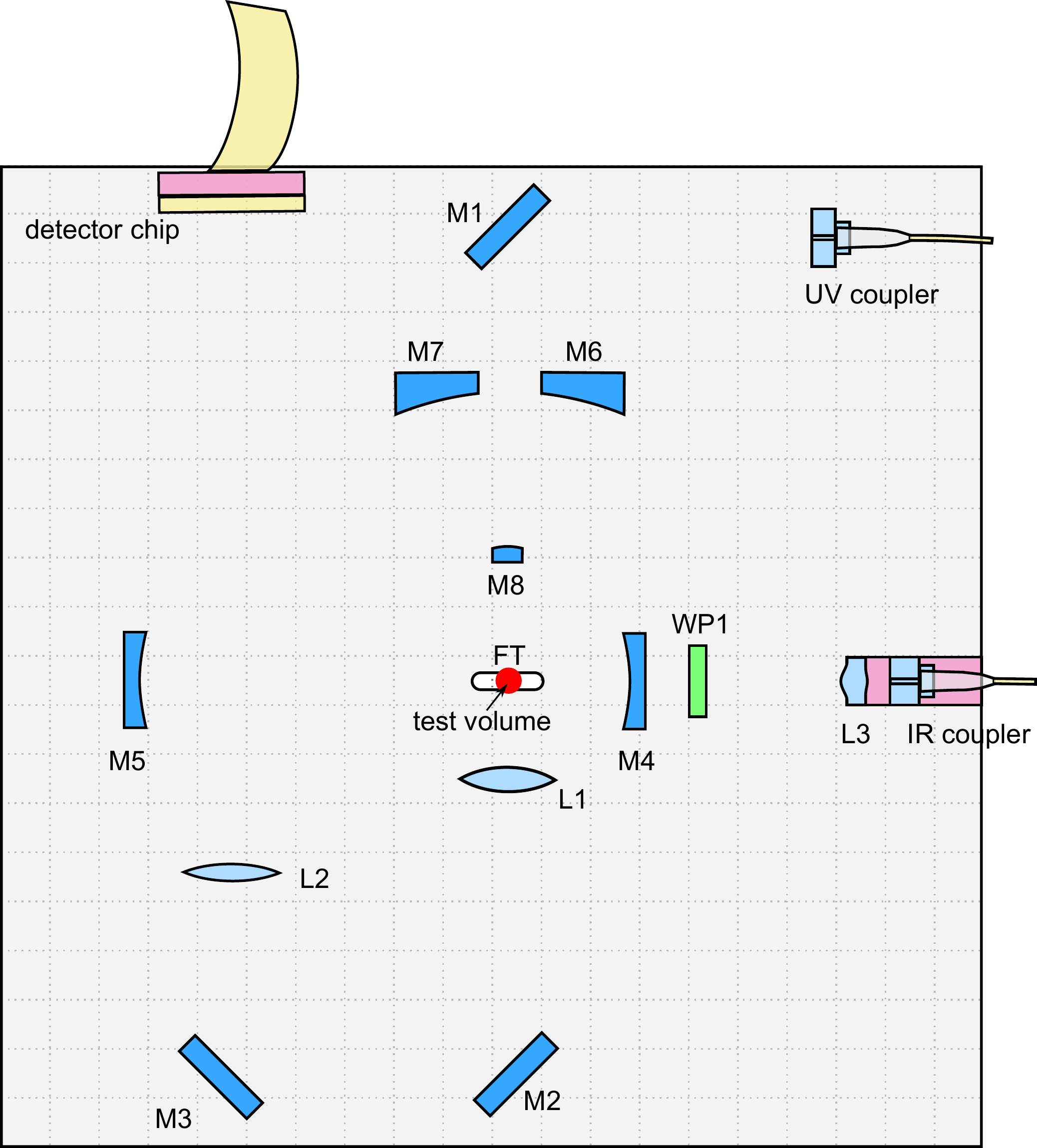}
  \includegraphics[width=0.45\linewidth]{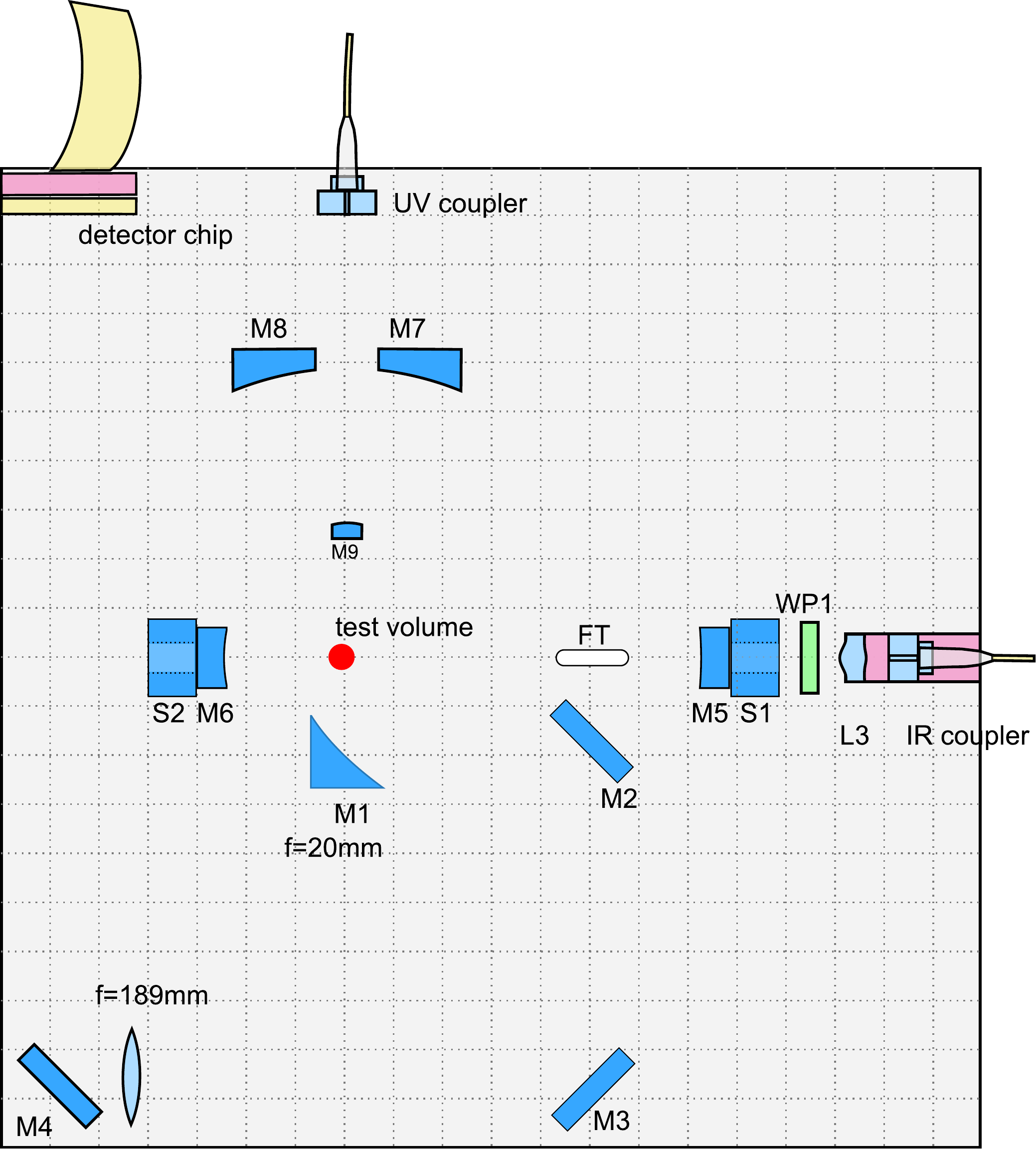}
  \caption{Preliminary design of the optical bench using refractive optics (left) and optimised design using reflective optics (right) for reducing the temperature of the test volume. M compoments represent mirrors, L components lenses, WP wave plate, S spacers and FT the slit where the nanospere is loaded in the cavity. \label{fig:optical-bench-layout}}
  \end{center}
\end{figure*}

\subsection{Optimizing the thermal properties of the optical bench \label{sub:Mirror vs Lens}}
For the present study, we optimised the optical bench for MAQRO with
respect to the temperature of the test volume by using reflective
rather than refractive optics for the imaging design. 

Figure \ref{fig:optical-bench-layout} schematically shows that a
combination of a parabolic mirror and several flat mirrors is now
used to image the trapping region onto a detector chip rather than
lenses. The camera to be used was developed for the James Webb
Space Telescope (JWST) \citep{Chip2,Loose2005a}. 

In this optimised design, the mirrors are bonded to a ZERODUR\textregistered{}
block with a central hole, which we refer to as spacer. Each spacer
is in turn bonded to the optical bench. A cavity design of this type
allows for maximum optical access to the nanosphere trapped in the
cavity mode and it is optimal for passive cooling of the test volume. 

Another design change is that the nanospheres are now loaded into
the cavity mode at a position approximately 24.25\,mm from the mirror
closest to the IR coupler, at $1/4$ of the cavity length.

In terms of the radiative influence of the various bench components
on the test volume, the most significant difference between both designs
is the substitution of the lens L1 with the parabolic mirror M1 near
the test volume. The advantage of the mirror is that it can be coated
with gold, which minimises the thermal radiation towards the test
volume. The reflecting area of the mirror is larger by a factor $\sqrt{2}$ than the original lens to keep the numerical aperture of the imaging system constant.

We calculated a temperature of 13.9\,K for the test volume considering
the preliminary bench with the lens L1. For the optimised bench using
reflective optics, this temperature decreases to 11.2\,K assuming
that the mirror M1 is finished with gold. However, the temperature
of the test volume temperature could rise to 14.3\,K if the mirror
M1 is not coated. This is higher than the temperature for the preliminary
design as a consequence of the adjustment of the mirror reflecting
area.

\subsection{Orbital cases }
We analysed the transient thermal behavior of the instrument for two
types of orbit: an orbit around the Lagrangian point L2 with respect to Sun \& Earth and a quasi-stationary
highly elliptical Earth orbit. The first orbit is evaluated using
a cool-down analysis of the whole instrument, whereas the latter intends
to indicate operational constraints of a possible space mission in
an Earth orbit.

The scenario with the orbit around L2 assumes that the whole
instrument has a starting temperature of $20\,^\circ$C and is
left to cool down by radiating to deep space. The orbit is considered
to be sun oriented if the normal vector to the spacecraft surface
points towards the sun throughout the orbit. This way the instrument
can be accommodated on the sun shaded part of the spacecraft and does
not receive any direct solar radiation. Because of the fact that the
L2 point lies on the line through the sun and the Earth beyond the
Earth at a distance of 1.5 million km, the optical bench is also completely
shielded against direct Earth radiation.

For the the highly elliptical Earth orbit, we choose the following
orbital parameters for the thermal analysis: altitude at apogee =
600\,000\,km, altitude at perigee = 600\,km, inclination $= 63.4\,^\circ$,
argument of periapsis = $0\,^\circ$; right ascension of the
ascending node = $0\,^\circ$. In this case the orbital period
is around 19.6 days. Using these values the instrument has a very
high view factor to the Earth at perigee and a low view factor to
the Earth at apogee.

We assumed an average sun radiation flux of 1369\,\-$\mathrm{W/m^{2}}$
at 1\,AU, an average Earth albedo reflectance of 0.3 and a uniform
Earth infrared emissivity of 1 at 257\,K. The criterion for a quasi-stationary
highly elliptical orbit is fulfilled if the temperatures of all nodes
must periodically recur at a predetermined orbital position after
successive cycles.

\subsubsection{Thermal-analysis results of the orbit around L2}

\begin{figure}[htbp]
 \begin{center}
  \includegraphics[width=0.97\linewidth]{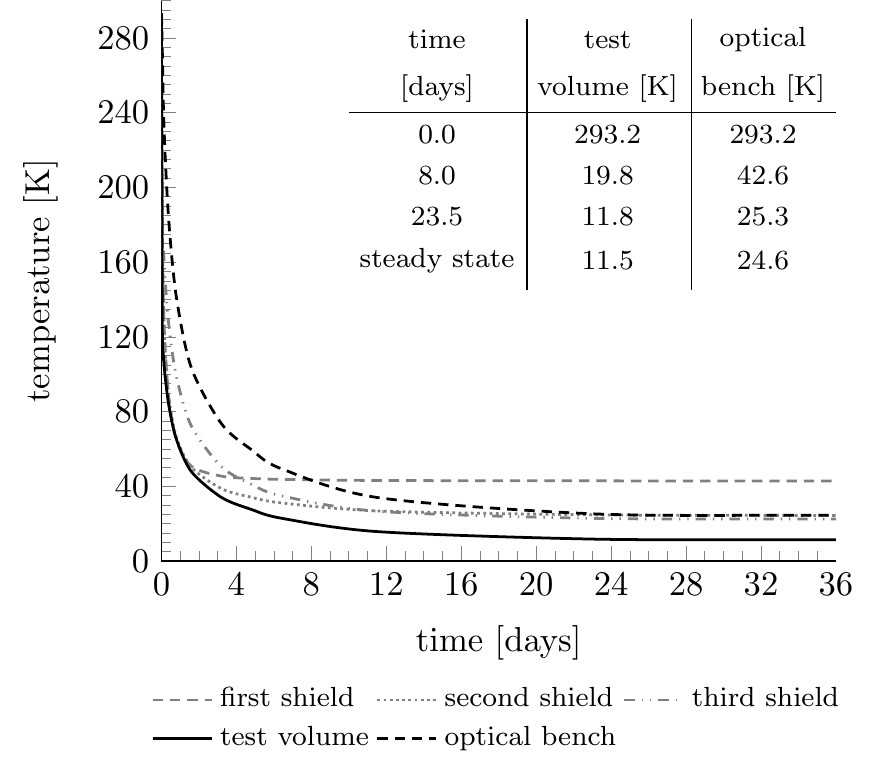}
  \caption{Thermal results of the scenario considering an orbit around the Sun-Earth Lagrangian point L2. \label{fig:Cooling-dynamics-L2}}
  \end{center}
\end{figure}

Figure \ref{fig:Cooling-dynamics-L2} shows the results of the thermal
analysis considering an L2 scenario. The evolution of the dropping
temperature of the passively cooled instrument has a fourth-degree
polynomial dependence on time. While the cooling progresses quickly
in the beginning, it then converges slowly towards the steady state
at later times. This is consistent with the fact that deep space acts
as the sole heat sink of the instrument. 

Among the instrument components investigated, the first shield is
the quickest to reach the steady state. It only takes around 3 days
because that component has the highest steady-state temperature compared
to other components. The last component to reach the steady state
is the optical bench as it has a low steady-state temperature and
a high heat capacity. It takes around 24 days for the bench to achieve
a temperature less than 25\,K.

The temperature of the test volume drops at the start of the cooling
because it has a high view factor to deep space and zero heat capacity.
It takes around 8 days for the test volume to cool down below 20\,K
and another 16 days for an additional temperature decrease of 8\,K.

\subsubsection{Thermal-analysis results of the quasi-stationary high\-ly elliptical
orbit}
\begin{figure}[htbp]
 \begin{center}
  \includegraphics[width=0.9\linewidth]{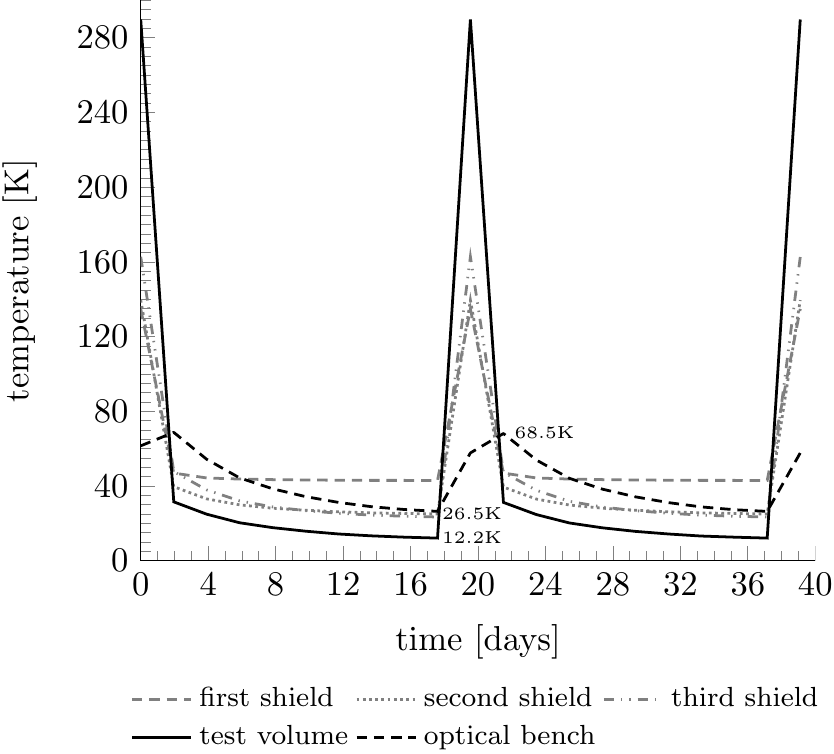}
  \caption{Thermal results of the scenario considering a quasi-stationary highly elliptical orbit. \label{fig:elliptic_results}}
  \end{center}
\end{figure}

Figure \ref{fig:elliptic_results} depicts the thermal behavior of
the instrument in a quasi-stationary highly elliptical orbit. The
temperature of an instrument component is the average of the temperatures
of its nodes. 

Our simulations showed that no solar heat flow was directly incident
on any node of the instrument. This confirms the sun orientation of
the spacecraft.

The temperatures of the instrument components reach their peaks right
after the perigee. After that, the instrument is cooled down while
traveling towards the apogee, where the Earth thermal influence diminishes
to a very large extent. The temperatures drop rapidly at first and
then slowly converge towards steady state during the cooling phase
before the next peak.

The temperature of the optical bench ranges between 26\,K and 69\,K
along the whole orbit. The maximum temperature change rate is around
10\,K/day. Because of the fact that the bench possesses a high heat
capacity at low temperatures, its temperature change is smaller than
for the other instrument components. 

It takes around seven days for the test volume to cool below 20\,K
after the point when it passes through perigee. Its temperature reaches
12.2\,K eleven days later. As the test volume is modelled as having
zero heat capacity, its temperature increases from 12.2\,K to 290\,K
within seven hours before reaching the perigee. However, this is also
the reason why its temperature decreases suddenly after leaving the
perigee.

\subsubsection{Discussion of the orbital cases}
The starting temperatures of the whole instrument for the L2 scenario
was $20\,^\circ$C. From this starting point, it takes about
eight days for the test volume to reach temperatures below the technical
requirements of MAQRO, i.e., below 20\,K. For the highly elliptical
orbit, we took the quasi-stationary temperatures in the instrument
at perigee as the starting condition. It then takes about seven days
for the test volume to cool down below 20\,K. The quantum experiments
must be interrupted each eleven days in this orbit due to the higher
heat loads at perigee. In contrast, the experiments can be performed
without interruptions for the L2 scenario.

In principle, a highly elliptical orbit scenario could offer more
flexibility in the mission planning as the instrument can be flown
on a satellite for Earth observation. Nevertheless, the bench is periodically
exposed to temperature changes due to the periodical heat loads, which
may cause, e.\,g., misalignment of the bench components, high
thermal stress\cite{Gwo2003a} or even structural
changes such as loss of the bonding force between interfaces due to
hysteresis effects caused by thermal cycles. In addition, the heat
load coming from the Earth is not uniformly distributed across the
instrument, which induces an uneven temperature distribution. 

Several concepts of shielding the instrument against Earth radiation
were analysed in order to reduce the effects of Earth radiation on
the instrument. These analyses showed that the measurements performed
by MAQRO are constrained to only 2 days even for the best-case scenario
investigated \citep{PilanZanoni2014}. The extra shields do not only
block a part of the Earth radiation at perigee, but also block a part
of the view factor of the instrument to deep space at apogee. Therefore,
the advantage gained by shielding the instrument against the hot radiation
from Earth is largely offset by the reduced efficiency of cooling
via thermal radiation to deep space and the additional thermal photons
received from the extra shields.

\subsubsection{Thermal results considering an SiC bench for the L2 scenario}

At the cryogenic temperatures expected for the optical bench in MAQRO,
ZERODUR\textregistered{} may not be the optimal choice of material.
In particular, at temperatures below 30\,K, Silicon Carbide (SiC)
has a significantly lower $CTE$ \citep{Roose2013a}. For that reason,
we also performed a thermal analysis for an optical bench made of
SiC in the case of an orbit around the Sun-Earth Lagrangian point L2. 

The results showed that the test volume cooled down significantly
faster with an SiC bench. It reached a temperature below 20\,K within
3 days (for the ZERODUR\textregistered{} bench: 8 days). The optical
bench reached a temperature of 25\,K in 5 days (for the ZERODUR\textregistered{}
bench: 24 days). This is due to the smaller heat capacity of SiC compared
to ZERODUR\textregistered{}. By extrapolating these results to the
case of a highly elliptical orbit, one can expect an extension of
the time period during which the thermal requirements of MAQRO are
fulfilled by around 5 days per orbital cycle \citep{PilanZanoni2014}.

Although SiC has a much higher thermal conductivity than ZERODUR\textregistered{},
the steady-state temperatures of the test volume for the ZERODUR\textregistered{}
bench (11.4\,K) and for the SiC bench (11.2\,K) are nearly identical.
This is consistent with the fact that the heat that reaches the optical
bench is rather low (see section \ref{sub:Heat-flux-analysis}).

\subsection{Further concept improvements }

Here, we discuss the potential for improvements in the passive-cooling
design in order to further optimise the st\-ea\-dy-state temperatures
of the test volume and the optical bench.

\subsubsection{Configuration of the MLI on the shields}

From the heat-flow diagram in Figure \ref{fig:Schematic-heat-flux},
one can see that the radiative heat transfer on the third shield is
comparatively low. Moreover, the oscillations of the spacecraft temperature
and the variations of the dissipation of the preprocessing chip are
strongly attenuated on the path from the oscillation source to the
MLI of the third shield (see Figure \ref{fig:Tannenbaum_Transferanalyse_CCD}).
For these reasons, we considered scenarios where the MLI was removed
on one or more shields. We present the corresponding results in Table
\ref{tab:Schilde-MLI}.

\begin{table}[htbp]
{\small{\caption{\label{tab:Schilde-MLI}Influence of the MLI of the shields on the
temperatures of the test volume, the bench and the third shield}
\medskip{}
}}{\small \par}

\begin{centering}
{\small{}}%
\begin{tabular}{ccccc}
\hline\hline
{\small{Analysis}} & & {\small{test}} & {\small{optical}} & {\small{$3^{\mathrm{rd}}$ shield}}\\
{\small{description}} & & {\small{volume}} & {\small{bench }} & plate\\
 & & {\small{ {[}K{]}}} & {\small{{[}K{]}}} & {\small{{[}K{]}}}\\
\hline 
{\small{Base configuration}} & & {\small{11.6}} & {\small{24.5}} & {\small{22.3}}\\
{\small{$3^{\mathrm{rd}}$ shield black painted }} & \multirow{2}{*}{\bigg{\}}} & \multirow{2}{*}{\small{11.6}} & \multirow{2}{*}{{\small{24.5}}} & \multirow{2}{*}{{\small{22.4}}}\\
{\small{and black MLI outer layer}} &  &  & \\ 
{\small{MLI removed}} & \multirow{2}{*}{\bigg{\}}} & \multirow{2}{*}{\small{11.6}} & \multirow{2}{*}{{\small{24.5}}} & \multirow{2}{*}{{\small{22.4}}}\\
{\small{from $3^{\mathrm{rd}}$ shield}} &  &  & \\
{\small{MLI removed}} & \multirow{2}{*}{\bigg{\}}} & \multirow{2}{*}{\small{11.8}} & \multirow{2}{*}{{\small{24.6}}} & \multirow{2}{*}{{\small{22.9}}}\\
{\small{from $2^{\mathrm{nd}}$ and $3^{\mathrm{rd}}$ shields}} &  &  & \\
{\small{MLI removed}} & \multirow{3}{*}{$\left.\rule{0cm}{0.6cm}\right\}$} & \multirow{3}{*}{\small{13.0}} & \multirow{3}{*}{{\small{27.5}}} & \multirow{3}{*}{{\small{26.4}}}\\
{\small{from $1^{\mathrm{st}}$, $2^{\mathrm{nd}}$}} &  &  & \\
{\small{and $3^{\mathrm{rd}}$ shields}} &  &  & \\
\hline\hline
\end{tabular}
\par\end{centering}{\small \par}

\end{table}

From these results, one can see that the presence or absence of the
MLI on the third shield effectively has no influence on the steady-state
temperatures of the test volume and the optical bench. Therefore,
the shield design can be simplified by removing the MLI from the third
shield. The aluminium plate of the third shield must be retained because
it is essential for the radiation of heat to deep space and acts as
a heat sink for the optical bench (see Figure \ref{fig:Schematic-heat-flux}).

Slight changes in the temperatures of the test volume and of the optical
bench were seen by additionally removing the MLI from the second shield.
Despite that, the MLI on the second shield should not be removed because
it has a significantly higher gain with respect to variations in the
dissipation heat of the preprocessing chip (0.021\,K/mW) than the
MLI on the third shield ($0.90\times10^{-3}\,\mathrm{K/mW}$).

Because of the fact that the heat flow on the MLI of the first shield
is comparatively high, its removal would cause a significant change
in the temperatures of the test volume and the optical bench. This
renders the MLI of the first shield essential for passive cooling.

\begin{table}[htbp]
\centering{}\caption{\label{tab:Elongated shields}Thermal results of the instrument with
elongated shields}
\medskip{}
{\small{}}%
\begin{tabular}{ccccccccc}
\hline\hline 
{\small{diameter of}} & \hspace{0.1cm} & {\small{MLI}} & \hspace{0.1cm} & {\small{test}} & \hspace{0.1cm} & {\small{optical }} & \hspace{0.1cm} & {\small{outer layer of}}\\
{\small{$1^{\mathrm{st}}$ shield}} & & {\small{outer}} & & {\small{volume}} & & {\small{bench }} & & {\small{$1^{\mathrm{st}}$shield MLI}}\\
{\small{{[}m{]}}} & & layer & & {\small{ {[}K{]}}} & & {\small{{[}K{]}}} & & {\small{{[}K{]}}}\\
\hline
{\small{0.9}} & & {\small{gold}} & & {\small{11.4}} & & \small{24.5} & & {\small{123.4}}\\
{\small{2.4}} & & {\small{gold}} & & {\small{9.7}} & & {\small{18.9}} & & {\small{520.5}}\\
{\small{2.4}} & & {\small{black}} & & {\small{9.7}} & & {\small{19.1}} & & {\small{370.7}}\\
\hline\hline
\end{tabular}
\end{table}

\subsubsection{Extension of the shields }

We showed in section \ref{sub:Heat-flux-analysis} that the shields
act as radiators in removing heat coming from the struts and from
the spacecraft to deep space. Is it possible to improve this effect
by using larger shields? We tried to address this issue by investigating
a design with extended shields. In particular, we analysed a design
where the shields are just large enough to still fit into a Soyuz-Fregat
fairing with a diameter of 2.8\,m. The configuration of the struts
and the dimensions of the interface with the spacecraft were the same
as described in the previous sections. In this modified configuration,
the first shield is extended beyond the spacecraft boundary and, therefore,
this shield receives direct solar radiation in a sun-oriented orbit.
We used the Monte Carlo method for modeling the solar radiation and
we present the results of this analysis in Table \ref{tab:Elongated shields}.

Our results show that extending the shields leads to a reduction of
the temperature of the test volume and that of the optical bench by
1.7\,K and 5.5\,K, respectively. In contrast to that, the temperature
of the outer layer of the MLI of the first shield increases by at
least 247\,K because of the exposure of the shield to solar radiation.

The outer layer of the MLI has a strong influence on its temperature
as a difference of 150\,K between the cases with gold and black Kapton
was observed. As $\alpha/\varepsilon_{\mathrm{Gold}}\approx 3.00$
and $\alpha/\varepsilon_{\mathrm{Kapton(EOL)}}\approx0.75$, the gold
layer absorbs much more sun radiation than it itself can radiate through
infrared light. Therefore, the outer layer of the MLI must not be
coated with gold in order to avoid thermal degradation.

\subsubsection{Heat-flow analysis for the optical bench}

\begin{table}[htbp]
\centering{}\caption{\label{tab:bench_gold}Sensitivity analysis on the coating of the
top surface of the optical bench}
\medskip{}
\begin{tabular}{ccccccccc}
\hline \hline 
gold & \hspace{0.2cm} & \multicolumn{3}{c}{temperature {[}K{]}} & \hspace{0.2cm} & \multicolumn{3}{c}{radiated heat {[}$\mathrm{\mu}$W{]}}\\
\cline{3-5}\cline{7-9}
 area & & test & & optical & & $\varepsilon=0.02$ & \hspace{0.2cm} & $\varepsilon=0.80$\\
{[}$\mathrm{cm^{2}}${]} & & volume & & bench & & gold area & & area\\
\hline 
0 & & 18.4 & & 23.1 & & - & & 447\\ 
73.5 & & 12.4 & & 23.3 & & 2.17 & & 385\\ 
204 & & 11.1 & & 23.7 & & 6.70 & & 242\\
400 & & 11.1 & & 24.4 & & 14.1 & & -\\
\hline\hline  
\end{tabular}
\end{table}

\begin{figure}[htbp]
\centering{}\includegraphics[width=0.8\columnwidth]{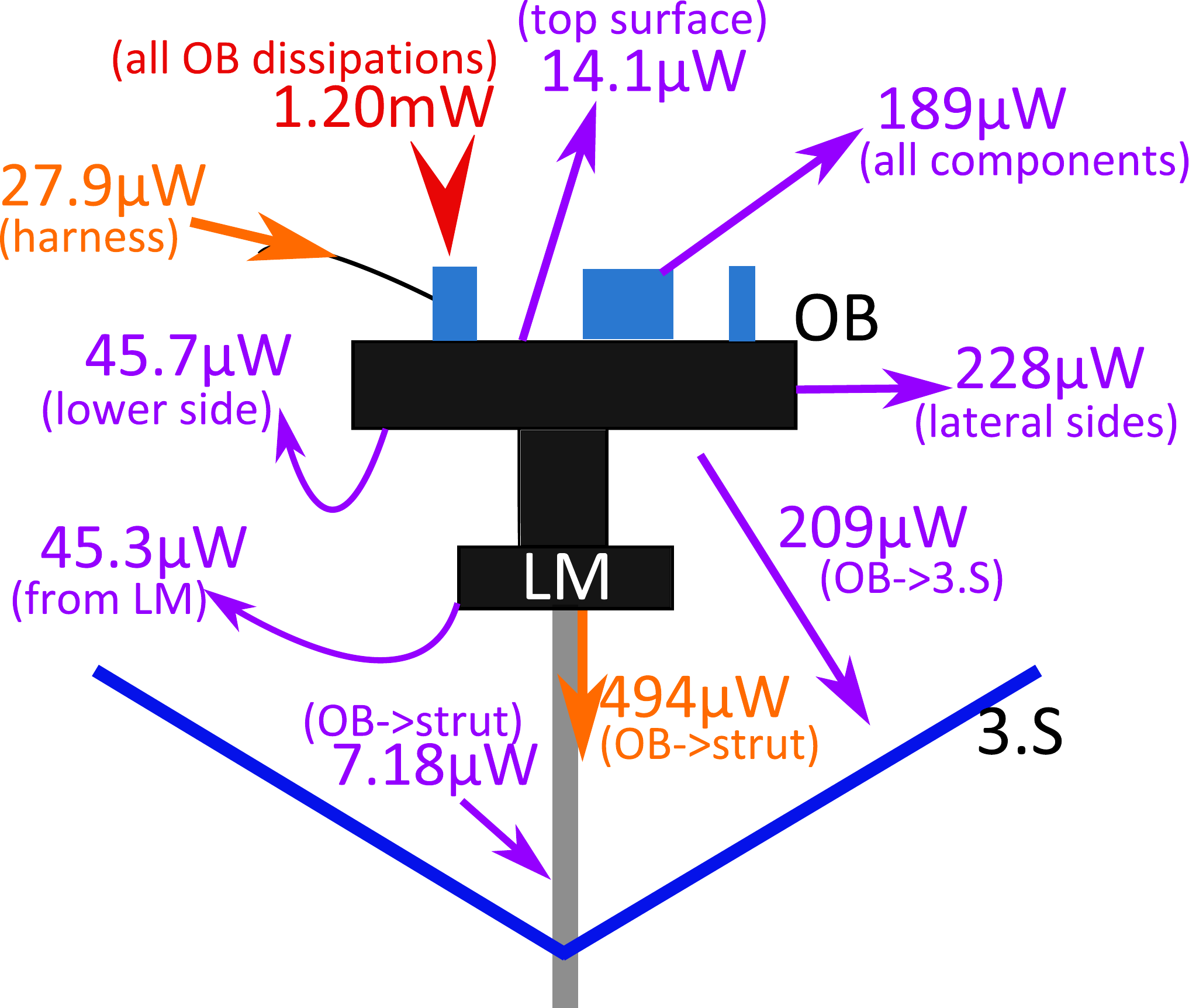}\caption{\label{fig:heatfluxbench}Schematic heat-flow diagram of the optical
bench}
\end{figure}

To probe for further possible design improvements, we analysed the
heat transfer in the optical bench The corresponding results are
shown in Figure \ref{fig:heatfluxbench}. To this end, we divided
the optical bench into its lower and lateral surfaces, the loading
mechanism (LM) and the optical components. The optical components
radiate a large part of heat to deep space because the dissipation
on the optical bench is largely caused by the detector chip and the
cavity mirrors. The top surface of the optical bench radiates to deep
space to a significantly smaller extent because of its gold finishing.

An investigation of the gold finishing of the top surface of the bench
was performed in order to maximise the radiation to deep space. This
was done by partially coating the upper surface. The area near the
test volume was finished with gold ($\varepsilon=0.02$), where the
test volume position is located above the center of this area. The
remaining area of the bench has $\varepsilon=0.80$. Table \ref{tab:bench_gold}
shows the results.

Increasing the gold area reduces the radiation of the top surface
of the bench to deep space. Consequently, the test volume receives
less radiation and its temperature is lower.

By increasing the area not coated with gold, the top surface of the
optical bench can radiate significantly more to deep space, reducing
the overall temperature of the optical bench. On the down side, this
radiation increases the temperature of the test volume. 

More detailed investigations of the technical requirements of MAQRO,
in particular the decohering effects of a non-isotropic distribution
of thermal radiation in the immediate environment of the nanosphere,
should allow determining the ideal ratio between the gold-coated and
uncoated areas of the optical bench in the future.

\section{Conclusions}

We investigated several aspects of the performance and the design
of a radiatively cooled system for quantum optomechanical experiments
in the context of the proposed future space mission MAQRO. This instrument
consists of an optical bench externally mounted to the spacecraft
surface with a set of struts and shielded against direct heat exchange
with the hot spacecraft and the sun. The optical bench is used for
performing quantum measurements at cryogenic temperatures and microgravity
conditions in deep space. A geometric and a thermal mathematical model
of the instrument are implemented with the aid of numerical tools
such as ESATAN-TMS and TransFAST.

A heat-flow analysis of the entire instrument shows that the shields
not only block the spacecraft radiation, but also act as radiators
in receiving the heat from the instrument through the struts and emitting
it to deep space. Our analysis shows that the configuration of
the instrument consisting of three shields is thermally well optimised.
It also showed that positioning the preprocessing chip for optical
imaging below the first shield diverted the resulting dissipation
heat into the shield rather than into the optical bench.

A transfer-function analysis showed that the shield str\-uct\-ure
strongly attenuates variations of the spacecraft temperature as well
as of the dissipation of the preprocessing chip. For example, a change
of the spacecraft temperature by 5\,K results only in a change of
less than 0.01\,K in the temperature of the optical bench. Considering
a fluctuation period of 100\,s, the fractional frequency instability
$\left|\Delta L/L\right|$ of the cavity on the optical bench of
MAQRO is around $1.6\times10^{-16}$, which is comparable to the best
values achieved in ground-based cavity designs.

By a simple modification of the imaging optics on the optical bench
(replacing a lens with a gold-coated, parabolic mirror) we achieved
a reduction from 13.9\,K to 11.2\,K for the steady-state temperature
of the test volume.

In an analysis of various orbital cases, we showed that it takes about
8 days for the test volume and 24 days for the optical bench to passively
cool down to less than 20\,K in an orbit around L2 considering an initial temperature of $20^\circ\,$C for the
entire instrument. These times can be reduced significantly by using
SiC instead of ZERODUR\textregistered{} for the material of the optical
bench. For a highly elliptical orbit with an orbital period of about
20 days, the time the instrument needs to cool down below the technical
requirements of MAQRO would significantly restrict the time for experiments
to 11 days for each period in the case of a ZERODUR\textregistered{}
bench. This limitation is more relaxed for an SiC bench.

A reduction from 11.4\,K to 9.7\,K in the steady-state temperature
of the test volume and from 24.5\,K to 19.1\,K in that of the optical
bench can be obtained by extending the shields.

\section*{Acknowledgements}
We acknowledge support from the Austrian Research Promotion Agency (FFG, project no. 3589434).

\bibliography{heatshield_v2}

%merlin.mbs apsrev4-1.bst 2010-07-25 4.21a (PWD, AO, DPC) hacked
%Control: key (0)
%Control: author (8) initials jnrlst
%Control: editor formatted (1) identically to author
%Control: production of article title (-1) disabled
%Control: page (0) single
%Control: year (1) truncated
%Control: production of eprint (0) enabled
\begin{thebibliography}{40}%
\makeatletter
\providecommand \@ifxundefined [1]{%
 \@ifx{#1\undefined}
}%
\providecommand \@ifnum [1]{%
 \ifnum #1\expandafter \@firstoftwo
 \else \expandafter \@secondoftwo
 \fi
}%
\providecommand \@ifx [1]{%
 \ifx #1\expandafter \@firstoftwo
 \else \expandafter \@secondoftwo
 \fi
}%
\providecommand \natexlab [1]{#1}%
\providecommand \enquote  [1]{``#1''}%
\providecommand \bibnamefont  [1]{#1}%
\providecommand \bibfnamefont [1]{#1}%
\providecommand \citenamefont [1]{#1}%
\providecommand \href@noop [0]{\@secondoftwo}%
\providecommand \href [0]{\begingroup \@sanitize@url \@href}%
\providecommand \@href[1]{\@@startlink{#1}\@@href}%
\providecommand \@@href[1]{\endgroup#1\@@endlink}%
\providecommand \@sanitize@url [0]{\catcode `\\12\catcode `\$12\catcode
  `\&12\catcode `\#12\catcode `\^12\catcode `\_12\catcode `\%12\relax}%
\providecommand \@@startlink[1]{}%
\providecommand \@@endlink[0]{}%
\providecommand \url  [0]{\begingroup\@sanitize@url \@url }%
\providecommand \@url [1]{\endgroup\@href {#1}{\urlprefix }}%
\providecommand \urlprefix  [0]{URL }%
\providecommand \Eprint [0]{\href }%
\providecommand \doibase [0]{http://dx.doi.org/}%
\providecommand \selectlanguage [0]{\@gobble}%
\providecommand \bibinfo  [0]{\@secondoftwo}%
\providecommand \bibfield  [0]{\@secondoftwo}%
\providecommand \translation [1]{[#1]}%
\providecommand \BibitemOpen [0]{}%
\providecommand \bibitemStop [0]{}%
\providecommand \bibitemNoStop [0]{.\EOS\space}%
\providecommand \EOS [0]{\spacefactor3000\relax}%
\providecommand \BibitemShut  [1]{\csname bibitem#1\endcsname}%
\let\auto@bib@innerbib\@empty
%</preamble>
\bibitem [{\citenamefont {Romero-Isart}\ \emph {et~al.}(2011)\citenamefont
  {Romero-Isart}, \citenamefont {Pflanzer}, \citenamefont {Blaser},
  \citenamefont {Kaltenbaek}, \citenamefont {Kiesel}, \citenamefont
  {Aspelmeyer},\ and\ \citenamefont {Cirac}}]{RomeroIsart2011b}%
  \BibitemOpen
  \bibfield  {author} {\bibinfo {author} {\bibfnamefont {O.}~\bibnamefont
  {Romero-Isart}}, \bibinfo {author} {\bibfnamefont {A.~C.}\ \bibnamefont
  {Pflanzer}}, \bibinfo {author} {\bibfnamefont {F.}~\bibnamefont {Blaser}},
  \bibinfo {author} {\bibfnamefont {R.}~\bibnamefont {Kaltenbaek}}, \bibinfo
  {author} {\bibfnamefont {N.}~\bibnamefont {Kiesel}}, \bibinfo {author}
  {\bibfnamefont {M.}~\bibnamefont {Aspelmeyer}}, \ and\ \bibinfo {author}
  {\bibfnamefont {J.~I.}\ \bibnamefont {Cirac}},\ }\href {\doibase
  10.1103/PhysRevLett.107.020405} {\bibfield  {journal} {\bibinfo  {journal}
  {Phys. Rev. Lett.}\ }\textbf {\bibinfo {volume} {107}},\ \bibinfo {pages}
  {020405} (\bibinfo {year} {2011})}\BibitemShut {NoStop}%
\bibitem [{\citenamefont {Kaltenbaek}\ \emph
  {et~al.}(2012{\natexlab{a}})\citenamefont {Kaltenbaek}, \citenamefont
  {Hechenblaikner}, \citenamefont {Kiesel}, \citenamefont {Romero-Isart},
  \citenamefont {Schwab}, \citenamefont {Johann},\ and\ \citenamefont
  {Aspelmeyer}}]{MAQROtechnical}%
  \BibitemOpen
  \bibfield  {author} {\bibinfo {author} {\bibfnamefont {R.}~\bibnamefont
  {Kaltenbaek}}, \bibinfo {author} {\bibfnamefont {G.}~\bibnamefont
  {Hechenblaikner}}, \bibinfo {author} {\bibfnamefont {N.}~\bibnamefont
  {Kiesel}}, \bibinfo {author} {\bibfnamefont {O.}~\bibnamefont
  {Romero-Isart}}, \bibinfo {author} {\bibfnamefont {K.~C.}\ \bibnamefont
  {Schwab}}, \bibinfo {author} {\bibfnamefont {U.}~\bibnamefont {Johann}}, \
  and\ \bibinfo {author} {\bibfnamefont {M.}~\bibnamefont {Aspelmeyer}},\
  }\href {\doibase 10.1007/s10686-012-9292-3} {\bibfield  {journal} {\bibinfo
  {journal} {Experimental Astronomy}\ }\textbf {\bibinfo {volume} {34}},\
  \bibinfo {pages} {123} (\bibinfo {year} {2012}{\natexlab{a}})}\BibitemShut
  {NoStop}%
\bibitem [{\citenamefont {Swinyard}\ and\ \citenamefont
  {Nakagawa}(2009)}]{swinyard2009space}%
  \BibitemOpen
  \bibfield  {author} {\bibinfo {author} {\bibfnamefont {B.}~\bibnamefont
  {Swinyard}}\ and\ \bibinfo {author} {\bibfnamefont {T.}~\bibnamefont
  {Nakagawa}},\ }\href {\doibase 10.1007/s10686-008-9090-0} {\bibfield
  {journal} {\bibinfo  {journal} {Experimental Astronomy}\ }\textbf {\bibinfo
  {volume} {23}},\ \bibinfo {pages} {193} (\bibinfo {year} {2009})}\BibitemShut
  {NoStop}%
\bibitem [{\citenamefont {Hawarden}\ \emph {et~al.}(1992)\citenamefont
  {Hawarden}, \citenamefont {Cummings}, \citenamefont {Telesco},\ and\
  \citenamefont {Jr}}]{TimHowardenPaper}%
  \BibitemOpen
  \bibfield  {author} {\bibinfo {author} {\bibfnamefont {T.~G.}\ \bibnamefont
  {Hawarden}}, \bibinfo {author} {\bibfnamefont {R.~O.}\ \bibnamefont
  {Cummings}}, \bibinfo {author} {\bibfnamefont {C.~M.}\ \bibnamefont
  {Telesco}}, \ and\ \bibinfo {author} {\bibfnamefont {H.~A.~T.}\ \bibnamefont
  {Jr}},\ }\href {\doibase 10.1007/BF00212480} {\bibfield  {journal} {\bibinfo
  {journal} {Space Science Review}\ }\textbf {\bibinfo {volume} {61}},\
  \bibinfo {pages} {113} (\bibinfo {year} {1992})}\BibitemShut {NoStop}%
\bibitem [{\citenamefont {Hawarden}\ \emph {et~al.}(1996)\citenamefont
  {Hawarden}, \citenamefont {Crane}, \citenamefont {Thronson~Jr}, \citenamefont
  {Penny}, \citenamefont {Orlowska},\ and\ \citenamefont
  {Bradshaw}}]{TimHowarden2}%
  \BibitemOpen
  \bibfield  {author} {\bibinfo {author} {\bibfnamefont {T.~G.}\ \bibnamefont
  {Hawarden}}, \bibinfo {author} {\bibfnamefont {R.}~\bibnamefont {Crane}},
  \bibinfo {author} {\bibfnamefont {H.~A.}\ \bibnamefont {Thronson~Jr}},
  \bibinfo {author} {\bibfnamefont {A.~J.}\ \bibnamefont {Penny}}, \bibinfo
  {author} {\bibfnamefont {A.~H.}\ \bibnamefont {Orlowska}}, \ and\ \bibinfo
  {author} {\bibfnamefont {T.~W.}\ \bibnamefont {Bradshaw}},\ }in\ \href@noop
  {} {\emph {\bibinfo {booktitle} {Infrared and Submillimeter Space Missions in
  the Coming Decade}}}\ (\bibinfo  {publisher} {Springer},\ \bibinfo {year}
  {1996})\ pp.\ \bibinfo {pages} {45--56}\BibitemShut {NoStop}%
\bibitem [{\citenamefont {Lightsey}\ \emph {et~al.}(2012)\citenamefont
  {Lightsey}, \citenamefont {Atkinson}, \citenamefont {Clampin},\ and\
  \citenamefont {Feinberg}}]{JamesWebb}%
  \BibitemOpen
  \bibfield  {author} {\bibinfo {author} {\bibfnamefont {P.~A.}\ \bibnamefont
  {Lightsey}}, \bibinfo {author} {\bibfnamefont {C.}~\bibnamefont {Atkinson}},
  \bibinfo {author} {\bibfnamefont {M.}~\bibnamefont {Clampin}}, \ and\
  \bibinfo {author} {\bibfnamefont {L.~D.}\ \bibnamefont {Feinberg}},\ }\href
  {\doibase 10.1117/1.OE.51.1.011003} {\bibfield  {journal} {\bibinfo
  {journal} {Optical Engineering}\ }\textbf {\bibinfo {volume} {51}},\ \bibinfo
  {pages} {011003} (\bibinfo {year} {2012})}\BibitemShut {NoStop}%
\bibitem [{\citenamefont {de~Bruijne}(2012)}]{de2012science}%
  \BibitemOpen
  \bibfield  {author} {\bibinfo {author} {\bibfnamefont {J.}~\bibnamefont
  {de~Bruijne}},\ }\href {\doibase 10.1007/s10509-012-1019-4} {\bibfield
  {journal} {\bibinfo  {journal} {Astrophysics and Space Science}\ }\textbf
  {\bibinfo {volume} {341}},\ \bibinfo {pages} {31} (\bibinfo {year}
  {2012})}\BibitemShut {NoStop}%
\bibitem [{\citenamefont {Urgoiti}\ and\ \citenamefont
  {Migliorero}(2005)}]{urgoiti2005mechanisms}%
  \BibitemOpen
  \bibfield  {author} {\bibinfo {author} {\bibfnamefont {E.}~\bibnamefont
  {Urgoiti}}\ and\ \bibinfo {author} {\bibfnamefont {G.}~\bibnamefont
  {Migliorero}},\ }in\ \href
  {http://www.esmats.eu/esmatspapers/pastpapers/pdfs/2005/urgoiti1.pdf} {\emph
  {\bibinfo {booktitle} {11th European Space Mechanisms and Tribology
  Symposium, ESMATS 2005}}},\ Vol.\ \bibinfo {volume} {591}\ (\bibinfo {year}
  {2005})\ pp.\ \bibinfo {pages} {163--170}\BibitemShut {NoStop}%
\bibitem [{\citenamefont {Pilbratt}\ \emph {et~al.}(2010)\citenamefont
  {Pilbratt}, \citenamefont {Riedinger}, \citenamefont {Passvogel},
  \citenamefont {Crone}, \citenamefont {Doyle}, \citenamefont {Gageur},
  \citenamefont {Heras}, \citenamefont {Jewell}, \citenamefont {Metcalfe},
  \citenamefont {Ott} \emph {et~al.}}]{pilbratt2010herschel}%
  \BibitemOpen
  \bibfield  {author} {\bibinfo {author} {\bibfnamefont {G.}~\bibnamefont
  {Pilbratt}}, \bibinfo {author} {\bibfnamefont {J.}~\bibnamefont {Riedinger}},
  \bibinfo {author} {\bibfnamefont {T.}~\bibnamefont {Passvogel}}, \bibinfo
  {author} {\bibfnamefont {G.}~\bibnamefont {Crone}}, \bibinfo {author}
  {\bibfnamefont {D.}~\bibnamefont {Doyle}}, \bibinfo {author} {\bibfnamefont
  {U.}~\bibnamefont {Gageur}}, \bibinfo {author} {\bibfnamefont
  {A.}~\bibnamefont {Heras}}, \bibinfo {author} {\bibfnamefont
  {C.}~\bibnamefont {Jewell}}, \bibinfo {author} {\bibfnamefont
  {L.}~\bibnamefont {Metcalfe}}, \bibinfo {author} {\bibfnamefont
  {S.}~\bibnamefont {Ott}},  \emph {et~al.},\ }\href {\doibase
  10.1051/0004-6361/201014759} {\bibfield  {journal} {\bibinfo  {journal}
  {Astronomy \& Astrophysics}\ }\textbf {\bibinfo {volume} {518}},\ \bibinfo
  {pages} {L1} (\bibinfo {year} {2010})}\BibitemShut {NoStop}%
\bibitem [{\citenamefont {Tauber}\ \emph {et~al.}(2010)\citenamefont {Tauber},
  \citenamefont {Mandolesi}, \citenamefont {Puget} \emph
  {et~al.}}]{tauber2010planck}%
  \BibitemOpen
  \bibfield  {author} {\bibinfo {author} {\bibfnamefont {J.~A.}\ \bibnamefont
  {Tauber}}, \bibinfo {author} {\bibfnamefont {N.}~\bibnamefont {Mandolesi}},
  \bibinfo {author} {\bibfnamefont {J.-L.}\ \bibnamefont {Puget}},  \emph
  {et~al.},\ }\href {\doibase 10.1051/0004-6361/200912983} {\bibfield
  {journal} {\bibinfo  {journal} {Astronomy \& Astrophysics}\ }\textbf
  {\bibinfo {volume} {520}},\ \bibinfo {pages} {A1} (\bibinfo {year}
  {2010})}\BibitemShut {NoStop}%
\bibitem [{\citenamefont {Oran}\ and\ \citenamefont
  {Naumann}(1977)}]{oran1977preliminary}%
  \BibitemOpen
  \bibfield  {author} {\bibinfo {author} {\bibfnamefont {W.}~\bibnamefont
  {Oran}}\ and\ \bibinfo {author} {\bibfnamefont {R.}~\bibnamefont {Naumann}},\
  }\href@noop {} {\bibfield  {journal} {\bibinfo  {journal} {NASA STI/Recon
  Technical Report N}\ }\textbf {\bibinfo {volume} {77}},\ \bibinfo {pages}
  {31222} (\bibinfo {year} {1977})}\BibitemShut {NoStop}%
\bibitem [{\citenamefont {Kaltenbaek}\ \emph
  {et~al.}(2012{\natexlab{b}})\citenamefont {Kaltenbaek}, \citenamefont
  {Hechenblaikner}, \citenamefont {Kiesel}, \citenamefont {Blaser},
  \citenamefont {Gr\"oblacher}, \citenamefont {Hofer}, \citenamefont {Vanner},
  \citenamefont {Schwab}, \citenamefont {Johann},\ and\ \citenamefont
  {Aspelmeyer}}]{TN3}%
  \BibitemOpen
  \bibfield  {author} {\bibinfo {author} {\bibfnamefont {R.}~\bibnamefont
  {Kaltenbaek}}, \bibinfo {author} {\bibfnamefont {G.}~\bibnamefont
  {Hechenblaikner}}, \bibinfo {author} {\bibfnamefont {N.}~\bibnamefont
  {Kiesel}}, \bibinfo {author} {\bibfnamefont {F.}~\bibnamefont {Blaser}},
  \bibinfo {author} {\bibfnamefont {S.}~\bibnamefont {Gr\"oblacher}}, \bibinfo
  {author} {\bibfnamefont {S.}~\bibnamefont {Hofer}}, \bibinfo {author}
  {\bibfnamefont {W.}~\bibnamefont {Vanner}, \bibfnamefont {M~R~Wieczorek}},
  \bibinfo {author} {\bibfnamefont {K.~C.}\ \bibnamefont {Schwab}}, \bibinfo
  {author} {\bibfnamefont {U.}~\bibnamefont {Johann}}, \ and\ \bibinfo {author}
  {\bibfnamefont {M.}~\bibnamefont {Aspelmeyer}},\ }\href@noop {} {\emph
  {\bibinfo {title} {Macroscopic quantum experiments in space using massive
  mechanical resonators}}},\ \bibinfo {type} {Tech. Rep.}\ (\bibinfo
  {institution} {Study conducted under contract with the European Space Agency,
  Po P5401000400},\ \bibinfo {year} {2011--2012})\BibitemShut {NoStop}%
\bibitem [{\citenamefont {Hechenblaikner}\ \emph {et~al.}(2014)\citenamefont
  {Hechenblaikner}, \citenamefont {Hufgard}, \citenamefont {Burkhardt},
  \citenamefont {Kiesel}, \citenamefont {Johann}, \citenamefont {Aspelmeyer},\
  and\ \citenamefont {Kaltenbaek}}]{MAQROpaper1}%
  \BibitemOpen
  \bibfield  {author} {\bibinfo {author} {\bibfnamefont {G.}~\bibnamefont
  {Hechenblaikner}}, \bibinfo {author} {\bibfnamefont {F.}~\bibnamefont
  {Hufgard}}, \bibinfo {author} {\bibfnamefont {J.}~\bibnamefont {Burkhardt}},
  \bibinfo {author} {\bibfnamefont {N.}~\bibnamefont {Kiesel}}, \bibinfo
  {author} {\bibfnamefont {U.}~\bibnamefont {Johann}}, \bibinfo {author}
  {\bibfnamefont {M.}~\bibnamefont {Aspelmeyer}}, \ and\ \bibinfo {author}
  {\bibfnamefont {R.}~\bibnamefont {Kaltenbaek}},\ }\href {\doibase
  10.1088/1367-2630/16/1/013058} {\bibfield  {journal} {\bibinfo  {journal}
  {New Journal of Physics}\ }\textbf {\bibinfo {volume} {16}},\ \bibinfo
  {pages} {013058} (\bibinfo {year} {2014})}\BibitemShut {NoStop}%
\bibitem [{\citenamefont {Chang}\ \emph {et~al.}(2010)\citenamefont {Chang},
  \citenamefont {Regal}, \citenamefont {Papp}, \citenamefont {Wilson},
  \citenamefont {Ye}, \citenamefont {Painter}, \citenamefont {Kimble},\ and\
  \citenamefont {Zoller}}]{Chang2010a}%
  \BibitemOpen
  \bibfield  {author} {\bibinfo {author} {\bibfnamefont {D.~E.}\ \bibnamefont
  {Chang}}, \bibinfo {author} {\bibfnamefont {C.~A.}\ \bibnamefont {Regal}},
  \bibinfo {author} {\bibfnamefont {S.~B.}\ \bibnamefont {Papp}}, \bibinfo
  {author} {\bibfnamefont {D.~J.}\ \bibnamefont {Wilson}}, \bibinfo {author}
  {\bibfnamefont {J.}~\bibnamefont {Ye}}, \bibinfo {author} {\bibfnamefont
  {O.}~\bibnamefont {Painter}}, \bibinfo {author} {\bibfnamefont {H.~J.}\
  \bibnamefont {Kimble}}, \ and\ \bibinfo {author} {\bibfnamefont
  {P.}~\bibnamefont {Zoller}},\ }\href {\doibase 10.1073/pnas.0912969107}
  {\bibfield  {journal} {\bibinfo  {journal} {Proceedings of the National
  Academy of Sciences of the United States of America}\ }\textbf {\bibinfo
  {volume} {107}},\ \bibinfo {pages} {1005} (\bibinfo {year}
  {2010})}\BibitemShut {NoStop}%
\bibitem [{\citenamefont {Romero-Isart}\ \emph {et~al.}(2010)\citenamefont
  {Romero-Isart}, \citenamefont {Juan}, \citenamefont {Quidant},\ and\
  \citenamefont {Cirac}}]{RomeroIsart2010a}%
  \BibitemOpen
  \bibfield  {author} {\bibinfo {author} {\bibfnamefont {O.}~\bibnamefont
  {Romero-Isart}}, \bibinfo {author} {\bibfnamefont {M.~L.}\ \bibnamefont
  {Juan}}, \bibinfo {author} {\bibfnamefont {R.}~\bibnamefont {Quidant}}, \
  and\ \bibinfo {author} {\bibfnamefont {J.~I.}\ \bibnamefont {Cirac}},\ }\href
  {\doibase 10.1088/1367-2630/12/3/033015} {\bibfield  {journal} {\bibinfo
  {journal} {New J. Phys.}\ }\textbf {\bibinfo {volume} {12}},\ \bibinfo
  {pages} {033015} (\bibinfo {year} {2010})}\BibitemShut {NoStop}%
\bibitem [{\citenamefont {Barker}\ and\ \citenamefont
  {Shneider}(2010)}]{Barker2010a}%
  \BibitemOpen
  \bibfield  {author} {\bibinfo {author} {\bibfnamefont {P.~F.}\ \bibnamefont
  {Barker}}\ and\ \bibinfo {author} {\bibfnamefont {M.~N.}\ \bibnamefont
  {Shneider}},\ }\href {\doibase 10.1103/PhysRevA.81.023826} {\bibfield
  {journal} {\bibinfo  {journal} {Phys. Rev. A}\ }\textbf {\bibinfo {volume}
  {81}},\ \bibinfo {pages} {023826} (\bibinfo {year} {2010})}\BibitemShut
  {NoStop}%
\bibitem [{\citenamefont {Bassi}\ \emph {et~al.}(2013)\citenamefont {Bassi},
  \citenamefont {Lochan}, \citenamefont {Satin}, \citenamefont {Singh},\ and\
  \citenamefont {Ulbricht}}]{Bassi2013a}%
  \BibitemOpen
  \bibfield  {author} {\bibinfo {author} {\bibfnamefont {A.}~\bibnamefont
  {Bassi}}, \bibinfo {author} {\bibfnamefont {K.}~\bibnamefont {Lochan}},
  \bibinfo {author} {\bibfnamefont {S.}~\bibnamefont {Satin}}, \bibinfo
  {author} {\bibfnamefont {T.~P.}\ \bibnamefont {Singh}}, \ and\ \bibinfo
  {author} {\bibfnamefont {H.}~\bibnamefont {Ulbricht}},\ }\href {\doibase
  10.1103/RevModPhys.85.471} {\bibfield  {journal} {\bibinfo  {journal} {Rev.
  Mod. Phys.}\ }\textbf {\bibinfo {volume} {85}},\ \bibinfo {pages} {471}
  (\bibinfo {year} {2013})}\BibitemShut {NoStop}%
\bibitem [{\citenamefont {Hackerm{\"u}ller}\ \emph {et~al.}(2004)\citenamefont
  {Hackerm{\"u}ller}, \citenamefont {Hornberger}, \citenamefont {Brezger},
  \citenamefont {Zeilinger},\ and\ \citenamefont {Arndt}}]{Hackermueller2004a}%
  \BibitemOpen
  \bibfield  {author} {\bibinfo {author} {\bibfnamefont {L.}~\bibnamefont
  {Hackerm{\"u}ller}}, \bibinfo {author} {\bibfnamefont {K.}~\bibnamefont
  {Hornberger}}, \bibinfo {author} {\bibfnamefont {B.}~\bibnamefont {Brezger}},
  \bibinfo {author} {\bibfnamefont {A.}~\bibnamefont {Zeilinger}}, \ and\
  \bibinfo {author} {\bibfnamefont {M.}~\bibnamefont {Arndt}},\ }\href
  {\doibase 10.1038/nature02276} {\bibfield  {journal} {\bibinfo  {journal}
  {Nature}\ }\textbf {\bibinfo {volume} {427}},\ \bibinfo {pages} {711}
  (\bibinfo {year} {2004})}\BibitemShut {NoStop}%
\bibitem [{\citenamefont {Hackerm{\"u}ller}\ \emph {et~al.}(2003)\citenamefont
  {Hackerm{\"u}ller}, \citenamefont {Hornberger}, \citenamefont {Brezger},
  \citenamefont {Zeilinger},\ and\ \citenamefont {Arndt}}]{Hackermueller2003a}%
  \BibitemOpen
  \bibfield  {author} {\bibinfo {author} {\bibfnamefont {L.}~\bibnamefont
  {Hackerm{\"u}ller}}, \bibinfo {author} {\bibfnamefont {K.}~\bibnamefont
  {Hornberger}}, \bibinfo {author} {\bibfnamefont {B.}~\bibnamefont {Brezger}},
  \bibinfo {author} {\bibfnamefont {A.}~\bibnamefont {Zeilinger}}, \ and\
  \bibinfo {author} {\bibfnamefont {M.}~\bibnamefont {Arndt}},\ }\href
  {\doibase 10.1007/s00340-003-1312-6} {\bibfield  {journal} {\bibinfo
  {journal} {{Appl. Phys. B}}\ }\textbf {\bibinfo {volume} {77}},\ \bibinfo
  {pages} {781} (\bibinfo {year} {2003})}\BibitemShut {NoStop}%
\bibitem [{\citenamefont {Romero-Isart}(2011)}]{RomeroIsart2011c}%
  \BibitemOpen
  \bibfield  {author} {\bibinfo {author} {\bibfnamefont {O.}~\bibnamefont
  {Romero-Isart}},\ }\href {\doibase
  http://dx.doi.org/10.1103/PhysRevA.84.052121} {\bibfield  {journal} {\bibinfo
   {journal} {Phys. Rev. A}\ }\textbf {\bibinfo {volume} {84}},\ \bibinfo
  {pages} {052121} (\bibinfo {year} {2011})}\BibitemShut {NoStop}%
\bibitem [{\citenamefont {{ITP Engines UK Ltd}}(2010)}]{ESATANmanual}%
  \BibitemOpen
  \bibfield  {author} {\bibinfo {author} {\bibnamefont {{ITP Engines UK
  Ltd}}},\ }\href@noop {} {\emph {\bibinfo {title} {{ESATAN-TMS thermal
  engineering manual}}}},\ \bibinfo {type} {Tech. Rep.}\ (\bibinfo {address}
  {available at www.esatan-tms.com},\ \bibinfo {year} {2010})\BibitemShut
  {NoStop}%
\bibitem [{\citenamefont {Love}\ \emph {et~al.}(2004)\citenamefont {Love},
  \citenamefont {Hoffman}, \citenamefont {Lum}, \citenamefont {Ando},
  \citenamefont {Ritchie}, \citenamefont {Therrien}, \citenamefont {Toth},\
  and\ \citenamefont {Holcombe}}]{Chip2}%
  \BibitemOpen
  \bibfield  {author} {\bibinfo {author} {\bibfnamefont {P.~J.}\ \bibnamefont
  {Love}}, \bibinfo {author} {\bibfnamefont {A.~W.}\ \bibnamefont {Hoffman}},
  \bibinfo {author} {\bibfnamefont {N.~A.}\ \bibnamefont {Lum}}, \bibinfo
  {author} {\bibfnamefont {K.~J.}\ \bibnamefont {Ando}}, \bibinfo {author}
  {\bibfnamefont {W.~D.}\ \bibnamefont {Ritchie}}, \bibinfo {author}
  {\bibfnamefont {N.~J.}\ \bibnamefont {Therrien}}, \bibinfo {author}
  {\bibfnamefont {A.~G.}\ \bibnamefont {Toth}}, \ and\ \bibinfo {author}
  {\bibfnamefont {R.~S.}\ \bibnamefont {Holcombe}},\ }\href {\doibase
  10.1117/12.555219} {\bibfield  {journal} {\bibinfo  {journal} {{Proc. of
  SPIE}}\ }\textbf {\bibinfo {volume} {5499}},\ \bibinfo {pages} {86} (\bibinfo
  {year} {2004})}\BibitemShut {NoStop}%
\bibitem [{\citenamefont {Kiesel}\ \emph {et~al.}(2013)\citenamefont {Kiesel},
  \citenamefont {Blaser}, \citenamefont {Delic}, \citenamefont {Grass},
  \citenamefont {Kaltenbaek},\ and\ \citenamefont
  {Aspelmeyer}}]{Lasercooling3}%
  \BibitemOpen
  \bibfield  {author} {\bibinfo {author} {\bibfnamefont {N.}~\bibnamefont
  {Kiesel}}, \bibinfo {author} {\bibfnamefont {F.}~\bibnamefont {Blaser}},
  \bibinfo {author} {\bibfnamefont {U.}~\bibnamefont {Delic}}, \bibinfo
  {author} {\bibfnamefont {D.}~\bibnamefont {Grass}}, \bibinfo {author}
  {\bibfnamefont {R.}~\bibnamefont {Kaltenbaek}}, \ and\ \bibinfo {author}
  {\bibfnamefont {M.}~\bibnamefont {Aspelmeyer}},\ }\href {\doibase
  10.1073/pnas.1309167110} {\bibfield  {journal} {\bibinfo  {journal}
  {Proceedings of the National Academy of Sciences of the United States of
  America}\ }\textbf {\bibinfo {volume} {110}},\ \bibinfo {pages} {14180}
  (\bibinfo {year} {2013})}\BibitemShut {NoStop}%
\bibitem [{\citenamefont {Li}\ \emph {et~al.}(2011)\citenamefont {Li},
  \citenamefont {Kheifets},\ and\ \citenamefont {Raizen}}]{Lasercooling5}%
  \BibitemOpen
  \bibfield  {author} {\bibinfo {author} {\bibfnamefont {T.}~\bibnamefont
  {Li}}, \bibinfo {author} {\bibfnamefont {S.}~\bibnamefont {Kheifets}}, \ and\
  \bibinfo {author} {\bibfnamefont {M.~G.}\ \bibnamefont {Raizen}},\ }\href
  {\doibase 10.1038/nphys1952} {\bibfield  {journal} {\bibinfo  {journal}
  {Nature Physics}\ }\textbf {\bibinfo {volume} {7}},\ \bibinfo {pages} {527}
  (\bibinfo {year} {2011})}\BibitemShut {NoStop}%
\bibitem [{\citenamefont {Horak}\ \emph {et~al.}(1997)\citenamefont {Horak},
  \citenamefont {Hechenblaikner}, \citenamefont {Gheri}, \citenamefont
  {Stecher},\ and\ \citenamefont {Ritsch}}]{Lasercooling1}%
  \BibitemOpen
  \bibfield  {author} {\bibinfo {author} {\bibfnamefont {P.}~\bibnamefont
  {Horak}}, \bibinfo {author} {\bibfnamefont {G.}~\bibnamefont
  {Hechenblaikner}}, \bibinfo {author} {\bibfnamefont {K.~M.}\ \bibnamefont
  {Gheri}}, \bibinfo {author} {\bibfnamefont {H.}~\bibnamefont {Stecher}}, \
  and\ \bibinfo {author} {\bibfnamefont {H.}~\bibnamefont {Ritsch}},\ }\href
  {\doibase 10.1103/PhysRevLett.79.4974} {\bibfield  {journal} {\bibinfo
  {journal} {Physical review letters}\ }\textbf {\bibinfo {volume} {79}},\
  \bibinfo {pages} {4974} (\bibinfo {year} {1997})}\BibitemShut {NoStop}%
\bibitem [{\citenamefont {Kaltenbaek}(2013)}]{Kaltenbaek2013}%
  \BibitemOpen
  \bibfield  {author} {\bibinfo {author} {\bibfnamefont {R.}~\bibnamefont
  {Kaltenbaek}},\ }in\ \href {\doibase 10.1117/12.2027051} {\emph {\bibinfo
  {booktitle} {{Optical Trapping and Optical Micromanipulation X}}}},\ Vol.\
  \bibinfo {volume} {8810}\ (\bibinfo {year} {2013})\ p.\ \bibinfo {pages}
  {88100B},\ \Eprint {http://arxiv.org/abs/1307.7021} {arXiv:1307.7021
  [quant-ph]} \BibitemShut {NoStop}%
\bibitem [{\citenamefont {McNamara}\ \emph {et~al.}(2008)\citenamefont
  {McNamara}, \citenamefont {Vitale}, \citenamefont {Danzmann},\ and\
  \citenamefont {on~behalf of~the LISA Pathfinder Science
  Working~Team}}]{LISAarticle1}%
  \BibitemOpen
  \bibfield  {author} {\bibinfo {author} {\bibfnamefont {P.}~\bibnamefont
  {McNamara}}, \bibinfo {author} {\bibfnamefont {S.}~\bibnamefont {Vitale}},
  \bibinfo {author} {\bibfnamefont {K.}~\bibnamefont {Danzmann}}, \ and\
  \bibinfo {author} {\bibnamefont {on~behalf of~the LISA Pathfinder Science
  Working~Team}},\ }\href {http://stacks.iop.org/0264-9381/25/i=11/a=114034}
  {\bibfield  {journal} {\bibinfo  {journal} {Classical and Quantum Gravity}\
  }\textbf {\bibinfo {volume} {25}},\ \bibinfo {pages} {114034} (\bibinfo
  {year} {2008})}\BibitemShut {NoStop}%
\bibitem [{\citenamefont {Elliffe}\ \emph {et~al.}(2005)\citenamefont
  {Elliffe}, \citenamefont {Bogenstahl}, \citenamefont {Deshpande},
  \citenamefont {Hough}, \citenamefont {Killow}, \citenamefont {Reid},
  \citenamefont {Robertson}, \citenamefont {Rowan}, \citenamefont {Ward},\ and\
  \citenamefont {Cagnoli}}]{Bonding2005}%
  \BibitemOpen
  \bibfield  {author} {\bibinfo {author} {\bibfnamefont {E.~J.}\ \bibnamefont
  {Elliffe}}, \bibinfo {author} {\bibfnamefont {J.}~\bibnamefont {Bogenstahl}},
  \bibinfo {author} {\bibfnamefont {A.}~\bibnamefont {Deshpande}}, \bibinfo
  {author} {\bibfnamefont {J.}~\bibnamefont {Hough}}, \bibinfo {author}
  {\bibfnamefont {C.}~\bibnamefont {Killow}}, \bibinfo {author} {\bibfnamefont
  {S.}~\bibnamefont {Reid}}, \bibinfo {author} {\bibfnamefont {D.}~\bibnamefont
  {Robertson}}, \bibinfo {author} {\bibfnamefont {S.}~\bibnamefont {Rowan}},
  \bibinfo {author} {\bibfnamefont {H.}~\bibnamefont {Ward}}, \ and\ \bibinfo
  {author} {\bibfnamefont {G.}~\bibnamefont {Cagnoli}},\ }\href {\doibase
  10.1088/0264-9381/22/10/018} {\bibfield  {journal} {\bibinfo  {journal}
  {Classical and quantum gravity}\ }\textbf {\bibinfo {volume} {22}},\ \bibinfo
  {pages} {257} (\bibinfo {year} {2005})}\BibitemShut {NoStop}%
\bibitem [{\citenamefont {Loose}\ \emph {et~al.}(2005)\citenamefont {Loose},
  \citenamefont {Beletic}, \citenamefont {Blackwell}, \citenamefont {Garnett},
  \citenamefont {Wong}, \citenamefont {Hall}, \citenamefont {Jacobson},
  \citenamefont {Rieke},\ and\ \citenamefont {Winters}}]{Loose2005a}%
  \BibitemOpen
  \bibfield  {author} {\bibinfo {author} {\bibfnamefont {M.}~\bibnamefont
  {Loose}}, \bibinfo {author} {\bibfnamefont {J.}~\bibnamefont {Beletic}},
  \bibinfo {author} {\bibfnamefont {J.}~\bibnamefont {Blackwell}}, \bibinfo
  {author} {\bibfnamefont {J.}~\bibnamefont {Garnett}}, \bibinfo {author}
  {\bibfnamefont {S.}~\bibnamefont {Wong}}, \bibinfo {author} {\bibfnamefont
  {D.}~\bibnamefont {Hall}}, \bibinfo {author} {\bibfnamefont {S.}~\bibnamefont
  {Jacobson}}, \bibinfo {author} {\bibfnamefont {M.}~\bibnamefont {Rieke}}, \
  and\ \bibinfo {author} {\bibfnamefont {G.}~\bibnamefont {Winters}},\ }in\
  \href {\doibase 10.1117/12.619638} {\emph {\bibinfo {booktitle} {Optics \&
  Photonics}}}\ (\bibinfo {organization} {International Society for Optics and
  Photonics},\ \bibinfo {year} {2005})\ pp.\ \bibinfo {pages}
  {59040V--59040V}\BibitemShut {NoStop}%
\bibitem [{\citenamefont {Messerschmid}\ and\ \citenamefont
  {Fasoulas}(2011)}]{Fasoulas2011}%
  \BibitemOpen
  \bibfield  {author} {\bibinfo {author} {\bibfnamefont {E.}~\bibnamefont
  {Messerschmid}}\ and\ \bibinfo {author} {\bibfnamefont {S.}~\bibnamefont
  {Fasoulas}},\ }in\ \href@noop {} {\emph {\bibinfo {booktitle}
  {Raumfahrtsysteme}}}\ (\bibinfo  {publisher} {Springer Berlin Heidelberg},\
  \bibinfo {year} {2011})\ pp.\ \bibinfo {pages} {333--370}\BibitemShut
  {NoStop}%
\bibitem [{\citenamefont {Crank}\ and\ \citenamefont
  {Nicolson}(1947)}]{Crank1947a}%
  \BibitemOpen
  \bibfield  {author} {\bibinfo {author} {\bibfnamefont {J.}~\bibnamefont
  {Crank}}\ and\ \bibinfo {author} {\bibfnamefont {P.}~\bibnamefont
  {Nicolson}},\ }\href {\doibase 10.1007/BF02127704} {\bibfield  {journal}
  {\bibinfo  {journal} {Math. Proc. Camb. Philos. Soc.}\ }\textbf {\bibinfo
  {volume} {43}},\ \bibinfo {pages} {50 } (\bibinfo {year} {1947})}\BibitemShut
  {NoStop}%
\bibitem [{\citenamefont {Shampine}\ and\ \citenamefont
  {Gear}(1979)}]{Shampine1979a}%
  \BibitemOpen
  \bibfield  {author} {\bibinfo {author} {\bibfnamefont {L.~F.}\ \bibnamefont
  {Shampine}}\ and\ \bibinfo {author} {\bibfnamefont {C.~W.}\ \bibnamefont
  {Gear}},\ }\href {\doibase 10.1137/1021001} {\bibfield  {journal} {\bibinfo
  {journal} {{SIAM Review}}\ }\textbf {\bibinfo {volume} {21}},\ \bibinfo
  {pages} {1} (\bibinfo {year} {1979})}\BibitemShut {NoStop}%
\bibitem [{\citenamefont {Altenburg}\ and\ \citenamefont
  {Burkhardt}(2008)}]{altenburg2008application}%
  \BibitemOpen
  \bibfield  {author} {\bibinfo {author} {\bibfnamefont {M.}~\bibnamefont
  {Altenburg}}\ and\ \bibinfo {author} {\bibfnamefont {J.}~\bibnamefont
  {Burkhardt}},\ }\href {\doibase 10.4271/2008-01-2076} {\emph {\bibinfo
  {title} {Application of Linear Control Methods to Satellite Thermal
  Analysis}}},\ \bibinfo {type} {Tech. Rep.}\ (\bibinfo  {institution} {ICES
  2008-01-2076},\ \bibinfo {address} {San Francisco},\ \bibinfo {year}
  {2008})\BibitemShut {NoStop}%
\bibitem [{\citenamefont {Altenburg}\ and\ \citenamefont
  {Burkhardt}(2010)}]{TransFAST2010}%
  \BibitemOpen
  \bibfield  {author} {\bibinfo {author} {\bibfnamefont {M.}~\bibnamefont
  {Altenburg}}\ and\ \bibinfo {author} {\bibfnamefont {J.}~\bibnamefont
  {Burkhardt}},\ }in\ \href {https://exchange.esa.int/thermal-workshop} {\emph
  {\bibinfo {booktitle} {{24th European Workshop on Thermal and ECLS
  Software}}}}\ (\bibinfo {address} {{ESTEC, Noordwijk, The Netherlands}},\
  \bibinfo {year} {2010})\BibitemShut {NoStop}%
\bibitem [{\citenamefont {Pilan~Zanoni}(2014)}]{PilanZanoni2014}%
  \BibitemOpen
  \bibfield  {author} {\bibinfo {author} {\bibfnamefont {A.}~\bibnamefont
  {Pilan~Zanoni}},\ }\emph {\bibinfo {title} {{Thermal Analysis and System
  Concept Improvements of the {MAQRO} Experiment for Measuring of Quantum
  Effects}}},\ \href@noop {} {Master's thesis},\ \bibinfo  {school} {TU Dresden
  and Airbus Defence {\&} Space} (\bibinfo {year} {2014})\BibitemShut {NoStop}%
\bibitem [{\citenamefont {Numata}\ \emph {et~al.}(2004)\citenamefont {Numata},
  \citenamefont {Kemery},\ and\ \citenamefont {Camp}}]{numata2004thermal}%
  \BibitemOpen
  \bibfield  {author} {\bibinfo {author} {\bibfnamefont {K.}~\bibnamefont
  {Numata}}, \bibinfo {author} {\bibfnamefont {A.}~\bibnamefont {Kemery}}, \
  and\ \bibinfo {author} {\bibfnamefont {J.}~\bibnamefont {Camp}},\ }\href
  {\doibase 10.1103/PhysRevLett.93.250602} {\bibfield  {journal} {\bibinfo
  {journal} {Physical Review Letters}\ }\textbf {\bibinfo {volume} {93}},\
  \bibinfo {pages} {250602} (\bibinfo {year} {2004})}\BibitemShut {NoStop}%
\bibitem [{\citenamefont {Notcutt}\ \emph {et~al.}(2006)\citenamefont
  {Notcutt}, \citenamefont {Ma}, \citenamefont {Ludlow}, \citenamefont
  {Foreman}, \citenamefont {Ye},\ and\ \citenamefont
  {Hall}}]{notcutt2006contribution}%
  \BibitemOpen
  \bibfield  {author} {\bibinfo {author} {\bibfnamefont {M.}~\bibnamefont
  {Notcutt}}, \bibinfo {author} {\bibfnamefont {L.-S.}\ \bibnamefont {Ma}},
  \bibinfo {author} {\bibfnamefont {A.~D.}\ \bibnamefont {Ludlow}}, \bibinfo
  {author} {\bibfnamefont {S.~M.}\ \bibnamefont {Foreman}}, \bibinfo {author}
  {\bibfnamefont {J.}~\bibnamefont {Ye}}, \ and\ \bibinfo {author}
  {\bibfnamefont {J.~L.}\ \bibnamefont {Hall}},\ }\href {\doibase
  10.1103/PhysRevA.73.031804} {\bibfield  {journal} {\bibinfo  {journal}
  {Physical Review A}\ }\textbf {\bibinfo {volume} {73}},\ \bibinfo {pages}
  {031804} (\bibinfo {year} {2006})}\BibitemShut {NoStop}%
\bibitem [{\citenamefont {Amairi}\ \emph {et~al.}(2013)\citenamefont {Amairi},
  \citenamefont {Legero}, \citenamefont {Kessler}, \citenamefont {Sterr},
  \citenamefont {W{\"u}bbena}, \citenamefont {Mandel},\ and\ \citenamefont
  {Schmidt}}]{amairi2013reducing}%
  \BibitemOpen
  \bibfield  {author} {\bibinfo {author} {\bibfnamefont {S.}~\bibnamefont
  {Amairi}}, \bibinfo {author} {\bibfnamefont {T.}~\bibnamefont {Legero}},
  \bibinfo {author} {\bibfnamefont {T.}~\bibnamefont {Kessler}}, \bibinfo
  {author} {\bibfnamefont {U.}~\bibnamefont {Sterr}}, \bibinfo {author}
  {\bibfnamefont {J.~B.}\ \bibnamefont {W{\"u}bbena}}, \bibinfo {author}
  {\bibfnamefont {O.}~\bibnamefont {Mandel}}, \ and\ \bibinfo {author}
  {\bibfnamefont {P.~O.}\ \bibnamefont {Schmidt}},\ }\href {\doibase
  10.1007/s00340-013-5464-8} {\bibfield  {journal} {\bibinfo  {journal}
  {Applied Physics B}\ }\textbf {\bibinfo {volume} {113}},\ \bibinfo {pages}
  {233} (\bibinfo {year} {2013})}\BibitemShut {NoStop}%
\bibitem [{\citenamefont {Gwo}(2003)}]{Gwo2003a}%
  \BibitemOpen
  \bibfield  {author} {\bibinfo {author} {\bibfnamefont {D.}~\bibnamefont
  {Gwo}},\ }\href {http://www.google.com/patents/US6548176} {\enquote {\bibinfo
  {title} {Hydroxide-catalyzed bonding},}\ } (\bibinfo {year} {2003}),\
  \bibinfo {note} {uS Patent 6,548,176}\BibitemShut {NoStop}%
\bibitem [{\citenamefont {Roose}\ and\ \citenamefont
  {Heltzel}(2013)}]{Roose2013a}%
  \BibitemOpen
  \bibfield  {author} {\bibinfo {author} {\bibfnamefont {S.}~\bibnamefont
  {Roose}}\ and\ \bibinfo {author} {\bibfnamefont {S.}~\bibnamefont
  {Heltzel}},\ }\href {\doibase 10.7795/810.20130822T} {\emph {\bibinfo {title}
  {High-precision measurements of the thermal expansion at cryogenic
  temperature on stable materials}}},\ \bibinfo {type} {Tech. Rep.}\ (\bibinfo
  {institution} {{Physikalisch-Technische Bundesanstalt (PTB)}},\ \bibinfo
  {year} {2013})\BibitemShut {NoStop}%
\end{thebibliography}%

\end{document}